\pdfoutput=1
\documentclass[11pt,a4paper]{article}

\usepackage{jheppub}
\usepackage{caption}
\usepackage{subcaption}
\usepackage{slashed}
	
\usepackage{pdfpages}

\usepackage{bm}
\usepackage{url}

\allowdisplaybreaks

\usepackage{ulem}%

\usepackage{feynmf} 
\def\Tr{\mbox{Tr}}

\def\al{\alpha}

\newcommand{\ep}{\varepsilon}

\def\be{\begin{equation}}
\def\ee{\end{equation}}
\def\bea{\begin{eqnarray}}
\def\eea{\end{eqnarray}}
\def\bse{\begin{subequations}}
\def\ese{\end{subequations}}
\def\bc{\begin{center}}
\def\ec{\end{center}}

\def\ra{\rightarrow}

\def\nonum{\nonumber}

\def\I{{\rm i}}

\def\D{{\rm d}}
\def\Ord{{\rm O}}

\def\Sp{{\slashed p}}

\newcommand{\ie}{{\it i.e.}}
\newcommand{\eg}{{\it e.g.}}

\begin{document}

\title{Two-loop mass anomalous dimension in reduced quantum electrodynamics and application to dynamical fermion mass generation}

\author{S.~Metayer and S.~Teber}

\affiliation{Sorbonne Universit\'e, CNRS, Laboratoire de Physique Th\'eorique et Hautes Energies, LPTHE, F-75005 Paris, France.}

\emailAdd{smetayer@lpthe.jussieu.fr}
\emailAdd{teber@lpthe.jussieu.fr}

\date{\today}

\abstract{
We consider reduced quantum electrodynamics (RQED$_{d_\gamma,d_e}$) a model describing fermions in a $d_e$-dimensional 
space-time and interacting via the exchange of massless bosons in $d_\gamma$-dimensions ($d_e \leq d_\gamma$). 
We compute the two-loop mass anomalous dimension, $\gamma_m$, in general RQED$_{4,d_e}$ with applications to RQED$_{4,3}$ and QED$_4$. 
We then proceed on studying dynamical (parity-even) fermion mass generation in RQED$_{4,d_e}$ by constructing a fully gauge-invariant gap equation 
for RQED$_{4,d_e}$ with $\gamma_m$ as the only input. This equation allows for a straightforward analytic computation of 
the gauge-invariant critical coupling constant, $\alpha_c$, which is such that a dynamical mass is generated for 
$\alpha_r > \alpha_c$, where $\alpha_r$ is the renormalized coupling constant, as well as the gauge-invariant critical number of fermion flavours, $N_c$, 
which is such that $\alpha_c \ra \infty$ and a dynamical mass is generated for $N < N_c$. For RQED$_{4,3}$, our results are in perfect agreement 
with the more elaborate analysis based on the resolution of truncated Schwinger-Dyson equations at two-loop order. In the case of QED$_4$, 
our analytical results (that use state of the art five-loop expression for $\gamma_m$) are in good quantitative agreement with those obtained from numerical approaches.

}

\maketitle

\begin{fmffile}{fmfrqedm}

\fmfset{dash_len}{2mm} 
\fmfset{arrow_len}{2.5mm} 
\fmfset{arrow_ang}{19} 
\def\vert#1{\fmfiv{decor.shape=circle,decor.filled=full,decor.size=2thick}{#1}}



\section{Introduction}
\label{sec.introduction}

Reduced quantum electrodynamics (RQED$_{d_\gamma,d_e}$) is a model describing relativistic fermions in a $d_e$-dimensional
space-time and interacting via the exchange of massless bosons in $d_\gamma$-dimensions ($d_e \leq d_\gamma$).
It has been introduced in \cite{Gorbar:2001qt} motivated by the study of dynamical chiral symmetry breaking in brane-world
theories (see also \cite{Kaplan:2009kr}). Soon after, a first application was devoted to the specially important 
case of conformal RQED$_{4,3}$ (also known as pseudo-QED from \cite{Marino:1992xi} and mixed-dimensional QED from the recent \cite{Son:2015xqa}) in relation with graphene 
\cite{Gorbar:2002iw}. More precisely, RQED$_{4,3}$ 
describes graphene \cite{Semenoff:1984dq} at its infra-red (IR) Lorentz invariant fixed point 
\cite{Gonzalez:1993uz} (see also \cite{Gusynin:2007ix,CastroNeto:2009zz,Kotov:2010yh,Teber:2018jdh} for reviews on graphene).  
It has been shown in \cite{Kotikov:2016yrn} that there is a mapping between 
RQED$_{4,3}$ and the well known three-dimensional QED (QED$_3$) in the large-$N$ limit
(where $N$ is the number of $4$-component Dirac spinors --- see also \cite{Gracey:2018ame} for a review on large-$N$ quantum field theories) 
which is a celebrated effective field theory for planar condensed matter physics systems exhibiting Dirac-like low-energy excitations such as, \eg, 
high-$T_c$ superconductors \cite{PhysRevB.42.4748,Dorey:1991kp,Franz:2001zz,Herbut:2002yq}, 
 and quantum antiferromagnets \cite{Farakos:1997hg}. 
The model of RQED$_{4,3}$ also captures some universal features of a broader range of so-called Dirac liquids that have been discovered experimentally 
during the last decade and are under active study such as, \eg, 
artificial graphene-like materials \cite{2013NatNa...8..625P}, 
surface states of topological insulators \cite{2010RvMP...82.3045H} 
and half-filled fractional quantum Hall systems \cite{2017NatPh..13.1168P}. 

Theoretically, there has been rather extensive studies of RQED$_{4,3}$ during the last decade with primary applications
to Dirac liquids, \eg, their transport and spectral properties \cite{Teber:2012de,Kotikov:2013kcl,Herbut:2013kca,Kotikov:2013eha,Valenzuela:2014uia,2016IJMPB..3050084H,Teber:2018goo}, quantum Hall effect \cite{2013arXiv1309.5879M,Son:2015xqa} (in \cite{Son:2015xqa} the model was invoked as an effective field theory describing half-filled fractional quantum Hall systems) 
and dynamical symmetry breaking \cite{Kotikov:2016yrn,2017arXiv170400381S} which will be our main focus in this paper (see also
\cite{Teber:2019kkp} for a review on these topics). From a more field theoretic point of view, the model was shown to be unitary 
\cite{Marino:2014oba}, it's properties were studied under the Landau-Khalatnikov-Fradkin transformation
\cite{Ahmad:2016dsb,James:2019ctc} as well as under duality transformations \cite{Hsiao:2017lch}. Renewed
interest in the model and its formal properties was triggered by the study  \cite{Herzog:2017xha} of 
interacting boundary conformal field theories, see, 
\eg,  \cite{Bashmakov:2017rko,Karch:2018uft,Dudal:2018pta,DiPietro:2019hqe,Giombi:2019enr} and supersymmetric extensions constructed and analyzed in \cite{Herzog:2018lqz,Gupta:2019qlg}. 
 
In this paper we shall focus on (parity-conserving) dynamical fermion mass generation in (massless) RQED$_{4,d_e}$ ($d_e \leq 4$) that include 
renormalizable models such as QED$_4$ and RQED$_{4,3}$ with dimensionless (renormalized) coupling constant $\al_r$ (the fine structure constant). Let's recall that 
this non-perturbative mechanism relies on the existence of a critical coupling constant, $\al_c$, which is such that
for $\al_r < \al_c$ fermions are massless (assuming chiral invariance), while for $\al_r > \al_c$ a dynamical
fermion mass is generated at a given number of fermion flavours $N$ (and chiral symmetry is dynamically broken). Alternatively, in the limit $\al_c \ra \infty$,
a dynamical mass is generated for $N < N_c$ where  $N_c$ is the critical number of fermion flavours. The knowledge of $\al_c$ and $N_c$
therefore provides precious information on the phase structure of gauge theories. 
 Moreover, in the condensed matter terminology, this mechanism corresponds to a (semi-)metal to insulator transition whereby a dynamical gap is generated at strong coupling.
It's study is crucial, \eg, for the development of graphene-based transistors \cite{2015PhRvL.115m6802N}.

In the last four decades, the standard  approaches to study dynamical mass generation in gauge theories were either based on lattice simulations
\cite{Dagotto:1988id,Dagotto:1989td,Hands:2004bh,Strouthos:2008kc,Karthik:2015sgq,Karthik:2016ppr}
or solving Schwinger-Dyson (SD) equations on which we shall focus here, see \cite{Roberts:1994dr} for an 
early detailed review as well as the manuscripts \cite{Bloch:1995dd,Reenders:1999fz}. 
While initial interests were towards four-dimensional theories, the importance of QED$_3$ (and its large-$N$ limit with no running coupling) 
was recognized very early \cite{Appelquist:1981vg,Appelquist:1981sf,Pisarski:1984dj} 
because of it's simpler UV structure\footnote{Because of the absence of running of their
coupling constants, large-$N$ QED$_3$ and RQED$_{4,3}$ may be referred to as ``standing'' gauge theories. In contrast, QED$_4$ has a running of the coupling constant. It is only in the quenched approximation (where fermion loops are neglected) that the coupling constant of QED$_4$ does not renormalize. } 
and similarity to quantum chromodynamics (QCD). Nevertheless, because of the formidable complexity of the task, 
calculations were generally carried out only at the leading order (LO) in the coupling 
constant \cite{Appelquist:1988sr} or by including non-perturbative ans\"atze for the vertex 
function in one-loop like SD equations \cite{Ball:1980ay,Curtis:1990zs,Kizilersu:2009kg}. 
Often, the resulting solution displayed residual gauge variance which is unsatisfactory for a physical quantity such as a critical coupling. 
Following early multi-loop works of Nash \cite{Nash:1989xx} and Kotikov \cite{Kotikov:1989nm,Kotikov:2011kg}, a complete
gauge-invariant prescription up to next-to-leading (NLO) of the $1/N$-expansion for QED$_3$ appeared only recently in 
\cite{Gusynin:2016som} and \cite{Kotikov:2016wrb,Kotikov:2016prf} (see also \cite{Kotikov:2020slw} 
for a recent review). In \cite{Kotikov:2016yrn} the results were  then mapped to RQED$_{4,3}$ thereby 
extending the LO results of \cite{Gorbar:2001qt} to the NLO in $\al$.     

The gauge-invariant prescriptions found in \cite{Gusynin:2016som} and \cite{Kotikov:2016prf} 
for large-$N$ QED$_3$ alleviate doubts about the validity of the SD equation approach though a similar prescription still has to be implemented for QED$_4$. 
Considering the complexity of the calculations,  simpler approaches are worthwhile investigating. 
An argument often invoked in the recent literature on QED$_3$,  
see, \eg, \cite{DiPietro:2015taa,Giombi:2015haa,Chester:2016ref,Janssen:2016nrm,Herbut:2016ide,Gukov:2016tnp,DiPietro:2017vsp,DiPietro:2017kcd,Benvenuti:2018cwd,Benvenuti:2019ujm} 
(see also \cite{Khachatryan:2019yzg} for a review), is the fact that a fermion quadrilinear operator becomes relevant at criticality
(this has actually been noted in the early literature on four-dimensional models, see, \eg, \cite{Bardeen:1985sm,Leung:1985sn,Miransky:1988gk,Leung:1989hw,Kondo:1992sq}).
The computation of the anomalous dimension of the corresponding composite operator allows then to derive a marginality crossing equation 
--- as referred to in \cite{Benvenuti:2018cwd,Benvenuti:2019ujm,Khachatryan:2019yzg} --- in order to extract the value of the critical coupling. 
Actually, as noticed in \cite{Gusynin:2016som}, the SD gap equation incorporates such a criterion in a rather simple and efficient 
way, \ie, via the mass anomalous dimension (the anomalous dimension of the fermion bilinear mass operator) which is a gauge-invariant quantity 
governing the ultra-violet (UV) asymptotic behaviour of the fermion propagator \cite{Lane:1974he,Politzer:1976tv,Miransky:1985wzx,Cohen:1988sq} (see also the textbook \cite{Miransky:1994vk}).

In the  present paper we will compute the mass anomalous dimension, $\gamma_m$, up to two loops, in a general theory of RQED$_{4,d_e}$ 
generalizing recent results in RQED$_{4,3}$ \cite{DiPietro:2019hqe} and recovering well known ones in QED$_4$ (see the lectures \cite{Grozin:2005yg}). 
Following \cite{Gusynin:2016som}, we then construct a gap equation that allows us to 
study the critical properties of RQED$_{4,d_e}$. In the case of RQED$_{4,3}$, we straightforwardly recover the results of 
\cite{Kotikov:2016yrn} for the NLO critical coupling and flavour number. As for QED$_4$, our approach is semi-phenomenological, but our results
are in good quantitative agreement with those obtained from numerical solutions of SD equations, see, \eg, \cite{Bloch:1995dd,Kizilersu:2014ela}.     

The paper is organized as follows. In Sec.~\ref{sec.model} we present the model, the perturbative setup as well as our renormalization conventions.
In Sec.~\ref{sec.PT}, we present the one and two-loop calculations of $\gamma_m$ in RQED$_{4,d_e}$.
The critical properties of the model are then analysed in Sec.~\ref{sec.DMG} where the gap equation is derived and
applications to RQED$_{4,3}$ and QED$_4$ (both in the quenched and unquenched cases) are provided. The conclusion is given in Sec.~\ref{sec.conclusion}. 
Some conventions and results related to the master integrals entering our calculations are provided in App.~\ref{App:masters} 
and  exact results for all the computed diagrams are presented in App.~\ref{App:exact}.


\section{Model and conventions}
\label{sec.model}

In Minkowski space, the RQED$_{d_\gamma,d_e}$ action \cite{Gorbar:2001qt,Teber:2012de,Kotikov:2013eha} including (in order to compute $\gamma_m$) a bare (parity-even) fermion mass reads:
\be
S = \int \D^{d_e} x\, \bar{\psi}_\sigma \left( \I \gamma^{\mu_e} D_{\mu_e} -m \right) \psi^\sigma
- \int \D^{d_\gamma} x\,\left[ \frac{1}{4}\,F^{\mu_\gamma \nu_\gamma}\,F_{\mu_\gamma \nu_\gamma}
+\frac{1}{2\xi}\left(\partial_{\mu_\gamma}A^{\mu_\gamma}\right)^2 \right]\, ,
\label{action}
\ee
where $\psi_\sigma$ are $N$ flavors of 4-component Dirac spinors ($\sigma=1, \cdots ,N$) in $d_e$ dimensions ($\mu_e=0,1, \cdots ,d_e-1$)
of mass $m$, 
 $D_\mu=\partial_\mu+\I e A_\mu$ is the covariant derivative and $\xi$ the gauge fixing parameter for the
$d_\gamma$-dimensional gauge field ($\mu_\gamma=0,1, \cdots ,d_\gamma-1$).
 In the following, we will use dimensional regularization and parameterize the dimensions as:
\be
d_\gamma=4-2\ep_\gamma, \qquad d_e=4-2\ep_e-2\ep_\gamma\, ,
\label{dims}
\ee
where $\ep_\gamma$ is the regulator which is such that $\ep_\gamma \ra 0$ for RQED$_{4,d_e}$ while $\ep_e = (d_\gamma - d_e)/2$.

From the action (\ref{action}), the Feynman rules read:
\begin{subequations}
\label{FeynmanRules}
\begin{alignat}{3}
S_0(p) ~~& = 
\parbox{3cm}{\centering
\begin{fmfgraph*}(20,2)
    \fmfleft{i}
    \fmfright{o}
    \fmf{fermion,label=$p$}{i,o}
\end{fmfgraph*}
}
&& = ~~\frac{\I}{\Sp - m}\, ,
\label{FeynmanRules:S0}
\\
\Gamma_0^\mu ~~& =
\parbox{3cm}{\centering 
\hspace{0.45cm}
\begin{fmfgraph*}(20,10)
    \fmfleft{i}
    \fmfright{o1,o2}
    \fmflabel{$\mu$}{i}
    \fmf{photon}{i,c}
    \fmf{fermion}{c,o1}
    \fmf{fermion}{o2,c}
\end{fmfgraph*}
}
&& = ~~-\I e \gamma^\mu,
\label{FeynmanRules:Gamma0}
\\
D_0^{\mu \nu}(p) ~~& = 
\parbox{3cm}{\centering
\begin{fmfgraph*}(20,2)
    \fmfleft{i}
    \fmfright{o}
    \fmflabel{$\mu$}{i}
    \fmflabel{$\nu$}{o}
    \fmf{photon,label=$p$}{i,o}
\end{fmfgraph*}
}
&& = ~~\frac{\I}{(4\pi)^{\ep_e}}\frac{\Gamma(1-\ep_e)}{(-p^2)^{1-\ep_e}}\,\left( g^{\mu \nu} - (1-\tilde{\xi})\,\frac{p^{\mu} p^{\nu}}{p^2}\right),
\label{FeynmanRules:D0}
\end{alignat}
\end{subequations}
where the photon propagator is the reduced ($d_e$-dimensional) one obtained by integrating out the bulk
$d_\gamma - d_e$ degrees of freedom and the corresponding reduced gauge-fixing parameter reads:
\be
\tilde{\xi} = \ep_e + (1-\ep_e)\, \xi\, .
\label{txi}
\ee

Upon turning on interactions, the dressed fermion propagator reads:
\be
S(p) = \frac{\I}{\Sp-m-\Sigma(p)}\, ,
\label{dressed:S}
\ee
where the fermion self-energy has the following parameterization appropriate to the massive case:
\be
\Sigma(p) = \Sp\,\Sigma_V(p^2) + m\,\Sigma_S(p^2) \, ,
\label{def:Sigma}
\ee
and its vector and scalar parts can be extracted with the help of
\be
\Sigma_V(p^2) = \frac{-1}{4\,(-p^2)}\,\Tr[ \Sp \, \Sigma(p)], \qquad
\Sigma_S(p^2) = \frac{1}{4\, m}\,\Tr \big[ \Sigma(p) \big]\, .
\label{def:SigmaV+SigmaS}
\ee
On the other hand, the dressed photon propagator can be expressed as:
\begin{subequations}
\label{dressed:D}
\begin{align} 
&\hspace{3cm} D^{\mu\nu}(p) = d_\perp(p^2)\,\left( g^{\mu\nu} - \frac{p^\mu p^\nu}{p^2} \right) + d_\parallel(p^2)\,\frac{p^\mu p^\nu}{p^2}\, ,
\\
&d_\perp(p^2) = \frac{d_0(p^2)}{1 - \I p^2\,d_0(p^2)\,\Pi(p^2)}, \quad d_\parallel(p^2) = \tilde{\xi}\,d_0(p^2), \quad 
d_0(p^2) = \frac{\I}{(4\pi)^{\ep_e}}\,\frac{\Gamma(1-\ep_e)}{(-p^2)^{1-\ep_e}}\, ,  
\label{dressed:D+}
\end{align}
\end{subequations}
where the photon self-energy is parameterized as:
\be
\Pi^{\mu \nu}(p) =  (p^2 g^{\mu \nu} - p^{\mu} p^{\nu})\, \Pi(p^2), \qquad
\Pi(p^2) = \frac{- \Pi^\mu_\mu(p)}{(d_e-1)\,(-p^2)}\, .
\label{def:Pi}
\ee
Anticipating the RPA analysis of sec.~\ref{subsec:effective-coupling}, let's note that, in the case of RQED$_{4,3}$, the transverse part of the photon propagator, \eqref{dressed:D+}, reads:
\be
d_\perp(p^2) = \frac{\I}{2\sqrt{-p^2}}\,\frac{1}{1 - \sqrt{-p^2}\,\Pi(p^2)/2}\, \qquad (\text{case of RQED$_{4,3}$})\, ,
\label{dressed:dperp:RQED43}
\ee
which can be usefully contrasted with the QED$_4$ case:
\be
d_\perp(p^2) = \frac{-\I}{p^2}\,\frac{1}{1 - \Pi(p^2)}\, \quad \qquad \qquad \qquad (\text{case of QED$_{4}$})\, .
\label{dressed:dperp:QED4}
\ee

With this perturbative setup, the renormalization constants are defined as:
\be
\psi = Z_\psi^{1/2}\,\psi_r \, , \quad A = Z_A^{1/2}\,A_r \, , \quad \xi = Z_\xi\, \xi_r \, ,
\quad \al = \mu^{2\ep_\gamma}\,Z_\al\,\al_r \, , \quad m = Z_m\, m_r\, ,
\label{RenormalizationCts}
\ee
and the following conventions will be used:
\be
\overline{\mu}^2 = 4\pi e^{-\gamma_E}\,\mu^2 \, , \quad
\alpha = \frac{e^2}{4\pi}, \quad
\bar{\alpha} = \frac{\alpha}{4\pi}\,,
\label{conventions}
\ee
where $\gamma_E$ is Euler's constant, $\overline{\mu}$ is the renormalization scale appropriate to the modified minimal subtraction ($\overline{\text{MS}}$)
scheme in which all our calculations will be performed and $\bar{\alpha}$ is the bare reduced coupling constant.
Notice that, from Ward identities, $Z_\al = Z_A^{-1}$ and $Z_\xi = Z_A$. Moreover, $Z_m$ (which is of special interest to us here) and $Z_\psi$
can be computed from the fermion self-energy parts with the help of
\be
 \frac{1+\Sigma_S}{1-\Sigma_V}\,Z_m = \text{finite}, \qquad  (1-\Sigma_V)\,Z_\psi = \text{finite} \, .
\label{def:Zpsi+Zm}
\ee
The corresponding anomalous dimensions and beta function 
 can then be calculated in a standard way with the help of
\be
\gamma_X = \frac{\D\log Z_X}{\D\log \mu} \quad (X = \{m,\psi,A\}), 
\qquad \beta=\frac{\D\, \bar{\alpha}_r}{\D\log\mu}= -2\ep_\gamma\,\bar{\alpha}_r + \gamma_A\,\bar{\alpha}_r \, .
\label{def:gammaX+beta}
\ee
Notice that, in the $\overline{\text{MS}}$-scheme, the renormalization constants are Laurent series in $\ep_\gamma$. They can be written as:
\be
Z_X = 1 + \delta Z_{X}, \qquad 
\delta Z_{X} = \sum_{l=1}^{\infty} \delta Z_{lX}\,\bar{\alpha}_r^l =  \sum_{l=1}^{\infty}\sum_{j=1}^l\,Z_{X}^{(l,j)}\,\frac{\bar{\alpha}_r^l}{\ep_\gamma^j} \qquad (X = \{m,\psi,A\})\, .
\label{param:Z}
\ee
Substituting (\ref{param:Z}) in (\ref{def:gammaX+beta}) yields:
\be
\gamma_X = \sum_{l=1}^\infty \gamma_{lX}\,\bar{\alpha}_r^l, \qquad \gamma_{lX} = -2l\,Z_{X}^{(l,1)} \qquad (X = \{m,\psi,A\})\, .
\label{gen:gammaX}
\ee

\section{Perturbative calculations}
\label{sec.PT}

\subsection{Preliminaries}

In this section we compute $\gamma_m$ in RQED$_{4,d_e}$. From (\ref{gen:gammaX}) and (\ref{def:Zpsi+Zm}), this requires
the knowledge of the fermion self-energies $\Sigma_V$ and $\Sigma_S$ which can be extracted from (\ref{def:SigmaV+SigmaS}). 
From \cite{Vladimirov:1979zm}, we recall that, in standard four-dimensional quantum field theories (that correspond to $\ep_e =0$ and $\ep_\gamma \ra 0$), 
the anomalous dimensions cannot depend on external momenta and masses and, in the $\overline{\text{MS}}$-scheme, on $\gamma_E$ and $\zeta_2$.
Hence, as an IR rearrangement \cite{Vladimirov:1979zm}, we will compute the self-energies $\Sigma_V$ and $\Sigma_S$ in the massless limit, see 
\cite{Kotikov:2018wxe} for a review on the corresponding techniques. 

Let's note that $\gamma_\psi$ was already computed in \cite{Kotikov:2013eha} for RQED$_{4,d_e}$
and we shall follow the notations of that paper. In particular, the following parameters will be useful: 
\begin{subequations}
\label{params-RQED4de}
\begin{flalign}
&L_p=\log\left(\frac{-p^2}{\bar{\mu}^2}\right)\,,\quad \bar{L}_p = L_p - \psi(2-\ep_e)+ \psi(1)\, , \quad
K_1 = \frac{\Gamma^3(1-\ep_e)\Gamma(\ep_e)}{\Gamma(2-2\ep_e)}\, ,
\label{params-RQED4de-Lp+K1} \\
&\overline{\Psi}_1 = \psi(1-\ep_e) - \psi(1)\, , \qquad
\overline{\Psi}_2 = \psi(\ep_e) - 2\psi(1-\ep_e) + 2\psi(2-2\ep_e) - \psi(1)\, ,
\label{params-RQED4de-Psis}
\end{flalign}
\end{subequations}
where $\psi(x)$ is the digamma function; see also App.~A of \cite{Kotikov:2013eha} for the expansion of the relevant master integrals. 

In the following, we will provide the $\ep_\gamma$-expansion of individual graphs contributing to  $\Sigma_V$ and $\Sigma_S$ self-energies
for arbitrary $\ep_e$ (exact expressions for individual graphs contributing to $\Sigma_S$ and $\Sigma_V$ can be found in App.~\ref{App:exact} of the
present paper).\footnote{The expansion of individual graphs contributing to $\Sigma_V$ in RQED$_{4,d_e}$ was
not explicit in \cite{Kotikov:2013eha}. For clarity and completeness, we provide such expressions here which can be
usefully compared with the corresponding ones for $\Sigma_S$.}
Combined with (\ref{def:Zpsi+Zm}) and (\ref{gen:gammaX}), this will
allow us to compute the renormalization constants and anomalous dimensions in RQED$_{4,d_e}$. 
Actually, from the one-loop expansion of the polarization operator, we will see that the limits $\ep_\gamma \ra 0$ 
and $\ep_e \ra 0$ do not commute (a fact already noticed in \cite{Kotikov:2013eha}).
In order to recover QED$_4$ from our general results, we will also present improved expressions for both $\gamma_\psi$ and $\gamma_m$
valid for RQED$_{4,d_e}$.

\subsection{One-loop calculations}

\begin{figure}[t]
    \centering
    \includegraphics[scale=1.0]{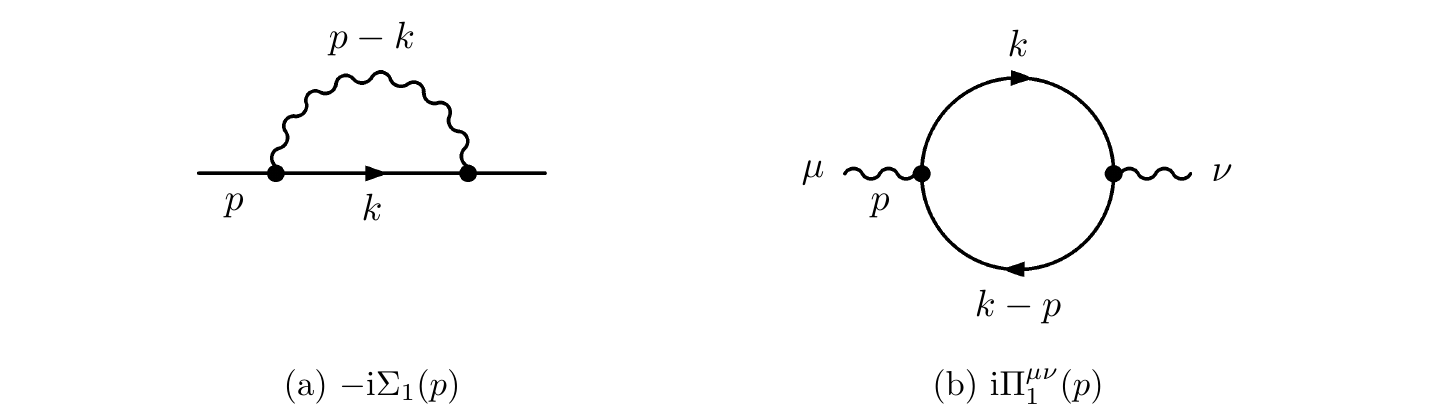}
    \captionsetup{width=.8\linewidth}
    \caption{1-loop fermion self energy and photon polarisation}
    \label{fig:fig1}
\end{figure}

At one-loop, we have the following two simple contributions to the fermion and photon self-energies:
\begin{subequations}
\label{one-loop:Pi+Sigma}
\begin{align}
-\I \Sigma_1(p) &= \mu^{2\ep_\gamma} \int [\D^{d_{e}} k]\, \Gamma_0^\mu\,S_0(k)\,\Gamma_0^\nu\,D^0_{\mu \nu}(p-k)\, ,
\label{one-loop:Sigma}
\\
\I \Pi_1^{\mu \nu}(p) &= -\mu^{2\ep_\gamma} N \int [\D^{d_{e}} k]\, \Tr \bigg[ \Gamma_0^\mu\,S_0(k)\,\Gamma_0^\nu\,S_0(k-p) \bigg]\, ,
\label{one-loop:Pi}
\end{align}
\end{subequations}
that are displayed in Fig.~\ref{fig:fig1}.

%
%

We first consider the one-loop fermion self-energy (\ref{one-loop:Sigma}). Using the parameterization (\ref{def:SigmaV+SigmaS}), 
we extract the scalar and vector parts that are then computed exactly (see \eqref{App:Sigma1+Pi1} of App.~\ref{App:exact}). The resulting 
$\ep_\gamma$-expansions at arbitrary $\ep_e$ then read:
\begin{subequations}
\label{Sigma1:V+S:exp}
\begin{flalign}
&\Sigma_{1S}(p^2) = \bar{\al}\,\left [ \left(\xi + \frac{3-\ep_e}{1-\ep_e} \right)\,\frac{1}{\ep_\gamma} +
\frac{1+\ep_e + (1-\ep_e)\xi }{(1-\ep_e)^2} - \left( \xi + \frac{3-\ep_e}{1-\ep_e} \right)\,\bar{L}_p \right .
\nonum \\
&\hspace{2.2cm}+ \bigg( \frac{1}{2}\,\left( \xi + \frac{3-\ep_e}{1-\ep_e} \right)\,\left( \bar{L}_p^2 + 2\zeta_2 - 3\,\psi'(1-\ep_e) \right)
- \frac{1+\ep_e + (1-\ep_e)\xi }{(1-\ep_e)^2}\,\bar{L}_p \bigg .
\nonum \\
&\hspace{2.2cm} \left . \bigg . + \frac{11 - \ep_e}{2(1 - \ep_e)^3}
+ \frac{5\xi}{2(1 - \ep_e)^2}
\bigg )\,\ep_\gamma + \Ord(\ep_\gamma^2) \right]\, ,
\label{Sigma1S:exp} \\
&\Sigma_{1V}(p^2) =  \bar{\al}\,\left [ \left( \frac{\ep_e}{2-\ep_e} - \xi \right)\,\frac{1}{\ep_\gamma} +
\frac{2\ep_e }{(2-\ep_e)^2} - \left( \frac{\ep_e}{2-\ep_e} - \xi \right)\,\bar{L}_p \right .
\nonum \\
&\hspace{2.2cm}+ \bigg( \frac{1}{2}\,\left(\frac{\ep_e}{2-\ep_e} - \xi \right)\,\left( \bar{L}_p^2 + 2\zeta_2 - 3\,\psi'(2-\ep_e) \right)
- \frac{2\ep_e}{(2-\ep_e)^2}\,\bar{L}_p \bigg .
\nonum \\
&\hspace{2.2cm}\left . \bigg . + \frac{4\ep_e}{(2-\ep_e)^3} \bigg)\,\ep_\gamma + \Ord(\ep_\gamma^2) \right]\, ,
\label{Sigma1V:exp} 
\end{flalign}
\end{subequations}
where $\bar{L}_p$ was defined in (\ref{params-RQED4de-Lp+K1}) and $\psi'(x)$ is the trigamma function. {For completeness, we also provide these expansions in the RQED$_{4,3}$ case ($\ep_\gamma \ra 0$ and $\ep_e=1/2$):
\begin{subequations}
\begin{flalign}
\Sigma_{1S}(p^2) & =\bar{\alpha} \left(\frac{\bar{\mu} ^2}{-4p^2}\right)^{\ep_\gamma }\!
\Bigg[\frac{5+\xi}{\ep_\gamma }+4 (4+\xi)
+\left(16 (4+\xi)-\frac{7\pi ^2}{12} (5+\xi)\right)\ep_\gamma
+\Ord\left(\ep_\gamma^2\right) \Bigg]\, ,
\\
\Sigma_{1V}(p^2) & =
\bar{\alpha}\left(
\frac{\bar{\mu} ^2}{-4p^2}\right)^{\ep_\gamma }\!
\Bigg[\frac{1-3\xi}{3\ep_\gamma }+\left(\frac{10}{9}-2 \xi \right)
+\left(\frac{112}{27}-8\xi-\frac{7\pi^2}{36}(1-3\xi)\right)\ep_\gamma
+\Ord\left(\ep_\gamma^2\right) \Bigg]\, .
\end{flalign}
\end{subequations}
}

With the help of \eqref{def:Zpsi+Zm}, these results allow 
for a straightforward derivation of the one-loop renormalization constants 
\begin{flalign}
\delta Z_{1m} = - \frac{2\,(3-2\ep_e)}{(1-\ep_e)(2-\ep_e)}\,\frac{1}{\ep_\gamma}, \qquad 
\delta Z_{1\psi} = \left( \frac{\ep_e}{2-\ep_e}  - \xi_r \right)\,\frac{1}{\ep_\gamma} \, ,
\label{Z1:psi+m} 
\end{flalign}
%
%
%
where the parameterization \eqref{param:Z} was used. From \eqref{gen:gammaX}, the corresponding one-loop anomalous dimensions read: 
\begin{flalign}
\gamma_{1m} = \frac{4(3-2\ep_e)}{(2-\ep_e)(1-\ep_e)}, \qquad
\gamma_{1\psi} = 2\,\left( \xi_r - \frac{\ep_e}{2-\ep_e} \right)\, .
\label{gamma1:psi+m}
\end{flalign}
%
%

Similarly, we now consider the one-loop polarization operator (\ref{one-loop:Pi}). Using the parameterization (\ref{def:Pi}) and from the exact result
\eqref{App:Pi1}, it's $\ep_\gamma$-expansion at arbitrary $\ep_e$ reads:
\be
\Pi_1(p^2)=-4\,\bar{\al}_r\,
N\,\frac{1 - \ep_e}{3 - 2\ep_e}\,\left(\frac{4\pi}{-p^2}\right)^{\ep_e}\,\frac{\Gamma^2(1-\ep_e)\,\Gamma(\ep_e)}{\Gamma(2-2\ep_e)}+\Ord(\ep_\gamma)\, ,
\label{Pi1:exp}
\ee
which is valid only for $d_e=3$. In this case ($\ep_e = 1/2$), it is finite:
\be
\Pi_1(p^2)= - \frac{\bar{\al}_r}{\sqrt{-p^2}}\,\hat{\Pi}_1^{(4,3)}, \qquad \hat{\Pi}_1^{(4,3)} = 2 N \pi^2\, ,
\label{Pi1:exp:RQED43}
\ee
which is related to the fact that the coupling constant does not run in RQED$_{4,3}$. 
However, (\ref{Pi1:exp}) does not reproduce the pole structure of QED$_4$. The correct result is obtained from (\ref{one-loop:Pi}) by first
setting $\ep_e = 0$ and then performing the $\ep_\gamma$-expansion; this yields:
\be
\Pi_1(p^2) = -\bar{\al}_r\left( \frac{4 N}{3}\,\left(\frac{1}{\ep_\gamma} - L_p\right) + \hat{\Pi}_1^{(4)} +\Ord(\ep_\gamma) \right), 
\qquad \hat{\Pi}_1^{(4)} = \frac{20 N}{9}\, .
\label{Pi1:exp:QED4}
\ee
This non-commutativity of the $\ep_e$ and $\ep_\gamma$ limits will appear in the two-loop fermion self-energy graphs with a fermion loop insertion
(see Fig.~\ref{fig:fig2}a) thereby affecting the anomalous dimensions. We will discuss improved expressions for them in Sec.~\ref{sec:AD}.

\subsection{Two-loop calculations}
\label{sec.two-loop}

\begin{figure}[t]
    \centering
    \includegraphics[scale=1.0]{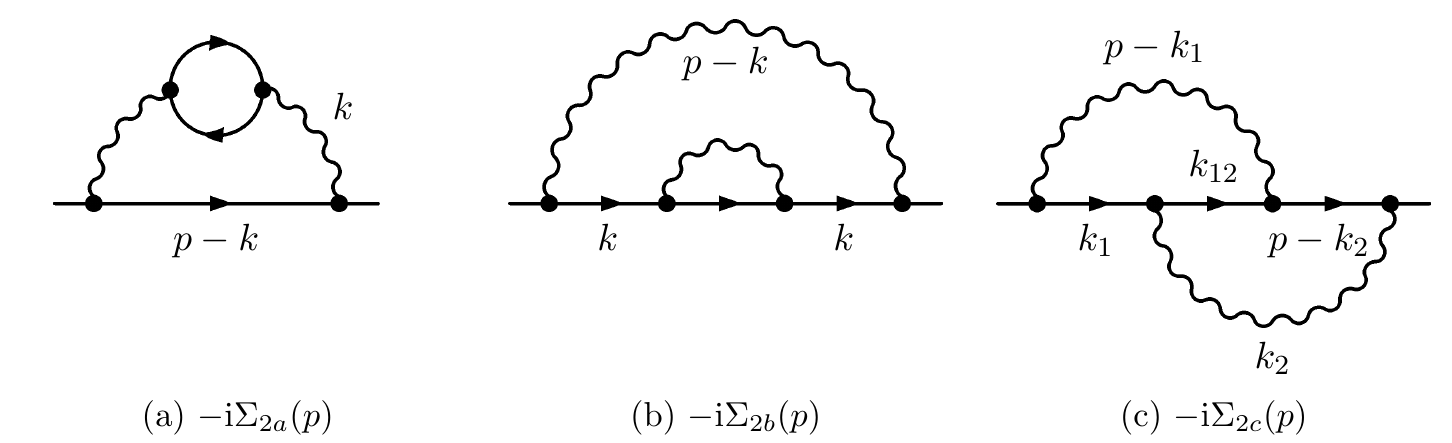}
    \captionsetup{width=.8\linewidth}
    \caption{2 loop contributions to the fermion self energy ($k_{12}=k_1-k_2$)}
    \label{fig:fig2}
\end{figure}

At two-loop, three diagrams contribute to the fermion self-energies:
\begin{subequations}
\begin{flalign}
-\I \Sigma_{2a}(p) &= \mu^{2\ep_\gamma}\int [\D^{d_e} k] \Gamma_0^\al\,S_0(p-k)\,\Gamma_0^\beta\,D_{0\,\al \mu}(k)\,\I \Pi_1^{\mu \nu}(k)\,D_{0\,\nu \beta}(k)\, ,
\\
-\I \Sigma_{2b}(p) &= \mu^{2\ep_\gamma}\int [\D^{d_e} k] \Gamma_0^\mu\,S_0(k)\,(-\I \Sigma_1(k))\,S_0(k)\,\Gamma_0^\nu\,D_{0\,\mu \nu}(p-k)\, ,
\\
-\I \Sigma_{2c}(p) &= \mu^{4\ep_\gamma}\int [\D^{d_e} k_1] [\D^{d_{e}} k_2] \Gamma_0^\mu\,S_0(p-k_2)\,D_{0\,\beta \mu}(k_2)\,\Gamma_0^\al\,S_0(k_{12})\,D_{0\,\al \nu}(p-k_1)\,\Gamma_0^\beta\,\,S_0(k_1)\,\Gamma_0^\nu\, ,
\end{flalign}
\end{subequations}
see Fig.~\ref{fig:fig2} where they are displayed.

Using the parameterization (\ref{def:SigmaV+SigmaS}), we extract the vector and scalar parts that are then computed exactly (see \eqref{App:Sigma2:S:a+b+c} and \eqref{App:Sigma2:V:a+b+c} of App.~\ref{App:exact}). 
For the scalar part, the resulting $\ep_\gamma$-expansions at arbitrary $\ep_e$ then read:
\begin{subequations}
\label{Sigma2:S:a+b+c:exp}
\begin{flalign}
&\Sigma_{2Sa}(p^2) = 2N\,\bar{\al}^2\,K_1\,\left [ -\frac{1}{\ep_\gamma} + 2\,\bar{L}_p + \overline{\Psi}_1 - \overline{\Psi}_2 + \Ord(\ep_\gamma) \right ]\, ,
\label{Sigma2Sa:exp} \\
&\Sigma_{2Sb}(p^2) =
\frac{\bar{\al}^2}{(2-\ep_e)\,(1-\ep_e)^2}\,\bigg [ \frac{\big(\ep_e - 3 - (1-\ep_e)\xi \big)\,\big(\ep_e\,(\ep_e + 3 - (3 - \ep_e)\,\xi) -6 + 2\xi  \big)}{2\,\ep_\gamma^2}    \bigg .
\nonum  \\
&+ \frac{1}{\ep_\gamma}\,\bigg( - \big( \ep_e - 3 - (1-\ep_e)\xi \big)\,\big( \ep_e\,(\ep_e + 3 - (3-\ep_e)\,\xi) - 6 + 2\xi \big)\,\bar{L}_p
\bigg .
\nonum \\
&+ \frac{1}{2\,(2-\ep_e)\,(1-\ep_e)}\,\bigg( 60  -  \ep_e\,\big( \ep_e\,(33 - 5\ep_e\,(6 - \ep_e)) + 40 \big)
+ 2\,\ep_e\,\big( \ep_e\,(7 + 3\ep_e\,(2-\ep_e)) - 22 \big)\,\xi \bigg .
\nonum \\
&\bigg . \bigg . + 24\xi - (2-\ep_e)^2\,(1-\ep_e)^2\,\xi^2 \bigg)  \bigg)
+
\big(\ep_e - 3 - (1-\ep_e)\xi \big)\,\big(\ep_e\,(\ep_e + 3 - (3 - \ep_e)\,\xi) -6 + 2\xi  \big)\,\bar{L}_p^2
\nonum \\
& - \frac{1}{(2-\ep_e)\,(1-\ep_e)}\,\bigg( 60  -  \ep_e\,\big( \ep_e\,(33 - 5\ep_e\,(6 - \ep_e)) + 40 \big)
+ 2\,\ep_e\,\big( \ep_e\,(7 + 3\ep_e\,(2-\ep_e)) - 22 \big)\,\xi \bigg .
\nonum \\
&\bigg .  + 24\xi - (2-\ep_e)^2\,(1-\ep_e)^2\,\xi^2 \bigg)\,\bar{L}_p + \big(\ep_e - 3 - (1-\ep_e)\xi \big)\,\big(\ep_e\,(\ep_e + 3 - (3 - \ep_e)\,\xi) -6 +
2\xi  \big)\,
\nonum \\
&\times \left( \frac{3}{2}\,\zeta_2 - 2\,\psi'(2-\ep_e) \right) + \frac{\ep_e\,\big(\ep_e\,(\ep_e\,(9\,\ep_e^2 - 84\,\ep_e +209) - 86) -268\big) + 248}{2\,(2-\ep_e)^2\,(1-\ep_e)^2}
\nonum \\
&\bigg. - \frac{\ep_e\,\big(\ep_e\,(\ep_e\,(11\,\ep_e - 35) + 10) +64 \big) -56}{(2-\ep_e)^2\,(1-\ep_e)}\,\xi
-\frac{2-\ep_e}{2}\,\xi^2 + \Ord(\ep_\gamma) \bigg ]\, ,
\label{Sigma2Sb:exp} \\
&\Sigma_{2Sc}(p^2) =
\frac{\bar{\al}^2}{(2-\ep_e)\,(1-\ep_e)}\,\left [ \frac{(\ep_e - (2- \ep_e)\,\xi)\,(\ep_e - 3 - (1 - \ep_e)\,\xi)}{\ep_\gamma^2} \right .
\nonum \\
&+
\frac{1}{\ep_\gamma}\,\bigg( -2\,(\ep_e - (2- \ep_e)\,\xi)\,(\ep_e - 3 - (1 - \ep_e)\,\xi)\,\bar{L}_p + \frac{5\,\ep_e}{2} -
\frac{\ep_e\,(\ep_e\,(\ep_e + 6) -23) + 18}{(2-\ep_e)\,(1-\ep_e)^2} \bigg .
\nonum \\
&+ \bigg. \left( \frac{2\,(2-\ep_e^2)}{(2-\ep_e)\,(1-\ep_e)} - 3\,\ep_e \right)\,\xi + \frac{3\,(2-\ep_e)}{2}\,\xi^2 \bigg)
+2\,(\ep_e - (2- \ep_e)\,\xi)\,(\ep_e - 3 - (1 - \ep_e)\,\xi)\,\bar{L}_p^2
\nonum \\
& - \bigg( 3\,(2-\ep_e)\,\xi^2 -
\left( 6\,\ep_e - \frac{8}{2-\ep_e} - \frac{4}{1-\ep_e} + 4 \right)\,\xi +5\,\ep_e - \frac{2\,(\ep_e\,(\ep_e\,(\ep_e + 6) - 23 ) + 18)}{(2-\ep_e)\,(1-\ep_e)^2} \bigg)\,\bar{L}_p
\nonum \\
& + \bigg( -3\,(2-\ep_e)\,(1-\ep_e)\,\xi^2 + 2\,(\ep_e\,(7-3\ep_e)-6)\xi - \frac{2}{1-\ep_e} - 16 + \ep_e\,(21-5\ep_e) \bigg)\,\psi'(1-\ep_e)
\nonum \\
& + 2\,\left( (2-\ep_e)\,(1-\ep_e)\,\xi^2 + (3-2\,\ep_e\,(2-\ep_e))\,\xi + 8 - \ep_e\,(9-2\ep_e) + \frac{1}{1-\ep_e} \right)\,\zeta_2
\nonum \\
& +  \frac{9\,(2-\ep_e)}{2\,(1-\ep_e)}\,\xi^2 - \frac{\ep_e\,(\ep_e\,(\ep_e\,(71 - 15\,\ep_e) - 134 ) + 128) -56}{(2-\ep_e)^2\,(1-\ep_e)^2}\,\xi
- \frac{92}{(2-\ep_e)^2\,(1-\ep_e)^3}
\nonum \\
& + \frac{\ep_e\,(\ep_e\,(\ep_e\,(\ep_e\,(28 + 3\,\ep_e) - 133) + 94) +172}{2\,(2-\ep_e)^2\,(1-\ep_e)^3}
\nonum \\
& \left . + 2\,(2-\ep_e)\,(1-\ep_e)\,(\ep_e\,(1-\ep_e)+1)\,\Gamma^2(1-\ep_e)\,G(4-2\ep_e,1-\ep_e,1,1-\ep_e,1,1) + \Ord(\ep_\gamma)
\right]\, ,
\label{Sigma2Sc:exp} 
\end{flalign}
\end{subequations}
where the parameters were defined in (\ref{params-RQED4de}). Notice that, the most complicated master integral $G(4-2\ep_e,1-\ep_e,1,1-\ep_e,1,1)$ 
only contributes to the finite part of the last graph, see \eqref{Sigma2Sc:exp}, and will therefore not affect the anomalous dimension. {In the RQED$_{4,3}$ case ($\ep_\gamma \ra 0$ and $\ep_e = 1/2$), these expansions simplify to:
\begin{subequations}
\begin{flalign}
\Sigma_{2Sa}(p^2) & = \bar{\alpha}^2 N\left(\frac{\bar{\mu}^2}{-p^2}\right)^{2\ep_\gamma}\!
\left[-\frac{2\pi^2}{\ep_\gamma}-8\pi^2+\Ord(\ep_\gamma)\right]\, ,
\\
\Sigma_{2Sb}(p^2) & = \bar{\alpha}^2 \left(\frac{\bar{\mu}^2}{-4p^2}\right)^{2\ep_\gamma}\!
\Bigg[\frac{(5+\xi)(17-3\xi)}{6\ep_\gamma^2}+\frac{1073+118\xi-27\xi^2}{9\ep_\gamma}
\nonum\\
& + \frac{2}{27}(11605+1550\xi-243\xi^2) - \frac{\pi^2}{4}(5+\xi)(17-3\xi) +\Ord(\ep_\gamma)\Bigg]\, ,
\\
\Sigma_{2Sc}(p^2) & = \bar{\alpha}^2 \left(\frac{\bar{\mu}^2}{-4p^2}\right)^{2\ep_\gamma}\!
\Bigg[-\frac{(5+\xi)(1-3\xi)}{3\ep_\gamma}-\frac{305-206\xi-63\xi^2}{9\ep_\gamma} -\frac{2}{27}(4507-1294\xi-513\xi^2)
\nonum \\
&  - \frac{\pi^2}{6}(27+22\xi+7\xi^2) +\frac{5\pi}{2}\, G(3,1/2,1,1/2,1,1) \Bigg]\, ,
\end{flalign}
\end{subequations}
where the non trivial $G(3,1/2,1,1/2,1,1)$, though not used in the following, is provided for completeness in \eqref{App:2-loop:complicatedG}. 
}

Proceeding in a similar way for the vector part, yields:
\begin{subequations}
\label{Sigma2:V:a+b+c:exp}
\begin{flalign}
&\Sigma_{2Va}(p^2) = 2N\,\bar{\al}^2\,K_1\,\left [ \frac{-\ep_e}{2-\ep_e}\,\frac{1}{\ep_\gamma} + 
\frac{2\ep_e }{2-\ep_e}\left( \bar{L}_p + \frac{1}{2}\,(\overline{\Psi}_1 - \overline{\Psi}_2) \right) + \frac{2}{1-\ep_e} - \frac{6}{(2-\ep_e)^2} 
+ \Ord(\ep_\gamma) \right]\, ,
\label{Sigma2Va:exp} \\
&\Sigma_{2Vb}(p^2) = \bar{\al}^2\,\left [  \frac{(\ep_e - (2-\ep_e)\xi)^2}{2(2-\ep_e)^2}\,\frac{1}{\ep_\gamma^2}    \right .
- \frac{1}{(2-\ep_e)^2}\,\bigg( (\ep_e - (2-\ep_e)\xi)^2\,\bar{L}_p  \bigg . 
\nonum \\
& \bigg . - \frac{(\ep_e - (2-\ep_e)\xi)(5\ep_e - (2-\ep_e)\xi)}{2(2-\ep_e)} \bigg)\,\frac{1}{\ep_\gamma} 
+ \frac{1}{(2-\ep_e)^2}\,\bigg ( (\ep_e - (2-\ep_e)\xi)^2\,\bar{L}_p^2 \bigg . 
\nonum \\
& -  
\left( \frac{5\ep_e^2}{2-\ep_e} - 6\ep_e\, \xi + (2-\ep_e)\xi^2 \right)\,\bar{L}_p 
+ \frac{3}{2}\,(\ep_e - (2-\ep_e)\xi)^2\,\zeta_2 + \frac{19\ep_e^2}{2(2-\ep_e)^2} - \frac{9\ep_e}{2-\ep_e}\, \xi 
\nonum \\
&\bigg . \bigg . - 2 \,(\ep_e - (2-\ep_e)\xi)^2\,\psi'(2-\ep_e)) + \frac{3\xi^2}{2} 
\bigg ) + \Ord(\ep_\gamma) \bigg ]\, , 
\label{Sigma2Vb:exp} \\
&\Sigma_{2Vc}(p^2) = 
- \bar{\al}^2\,\left [  \frac{(\ep_e - (2-\ep_e)\xi)^2}{(2-\ep_e)^2}\,\frac{1}{\ep_\gamma^2}    \right .
- \frac{1}{(2-\ep_e)^2}\,\bigg( 2\,(\ep_e - (2-\ep_e)\xi)^2\,\bar{L}_p  \bigg . 
\nonum \\
& \bigg . + \frac{\ep_e}{2}\,(\xi^2 + 10\xi + 1) - \xi^2 + 15 - \frac{20}{2-\ep_e} - \frac{2}{1-\ep_e}  \bigg)\,\frac{1}{\ep_\gamma} 
+ \frac{1}{(2-\ep_e)^2}\,\bigg ( 2\,(\ep_e - (2-\ep_e)\xi)^2\,\bar{L}_p^2 \bigg . 
\nonum \\
& +  
\big( \ep_e\,(\xi^2 +10 \xi +1) - 2\xi^2 + 30 - \frac{40}{2-\ep_e} - \frac{4}{1-\ep_e} \big)\,\bar{L}_p 
\nonum \\
& + 2\,\big(\ep_e^2\,(1+\xi)^2 - 4\,\ep_e\,(\xi^2+\xi+1) + 4\xi^2 + 6 \big)\,\zeta_2 + \frac{3\,(3\ep_e^2 - 10\ep_e + 9)}{2(1-\ep_e)^2}\, \xi^2 
\nonum \\
&- \left( \frac{26}{2-\ep_e} + \frac{6}{(1-\ep_e)^2} - 19 \right)\,\xi + \frac{\ep_e\,\big(\ep_e\,(\ep_e\,(13\ep_e+42) - 137) + 64 \big) + 28}{2\,(2-\ep_e)^2\,(1-\ep_e)^2} 
\nonum \\
&\bigg . \bigg . - \big( 3 \,(2-\ep_e)^2\,\xi^2 - 6\ep_e(2-\ep_e)\,\xi - (\ep_e\,(8-3\ep_e) - 12) \big)\,\psi'(1-\ep_e))
\bigg ) + \Ord(\ep_\gamma) \bigg ]\, , 
\label{Sigma2Vc:exp}
\end{flalign}
\end{subequations}
where the most complicated master integral $G(4-2\ep_e,1-\ep_e,1,1-\ep_e,1,1)$ only contributes to the $\Ord(\ep_\gamma)$ terms of \eqref{Sigma2Vc:exp} and is not even displayed in this equation. {In the RQED$_{4,3}$ case ($\ep_\gamma \ra 0$ and $\ep_e = 1/2$), these expansions simplify to:
\begin{subequations}
\begin{flalign}
\Sigma_{2Va}(p^2) & = \bar{\alpha}^2 N \left(\frac{\bar{\mu}^2}{-p^2}\right)^{2\ep_\gamma}\!
\left[-\frac{2\pi^2}{3\ep_\gamma}+\Ord(\ep_\gamma)\right]\, ,
\\
\Sigma_{2Vb}(p^2) & = \bar{\alpha}^2 \left(\frac{\bar{\mu}^2}{-4p^2}\right)^{2\ep_\gamma}\!
\Bigg[\frac{(1-3\xi)^2}{18\ep_\gamma^2}+\frac{(1-3\xi)(11-21\xi)}{27\ep_\gamma}+\frac{206}{81}-2\xi(6-7\xi)
\nonum \\
& -\frac{\pi^2}{12}(1-3\xi)^2+\Ord(\ep_\gamma)\Bigg]\, ,
\\
\Sigma_{2Vc}(p^2) & = \bar{\alpha}^2 \left(\frac{\bar{\mu}^2}{-4p^2}\right)^{2\ep_\gamma}\!
\Bigg[-\frac{(1-3\xi)^2}{9\ep_\gamma^2}-\frac{37-3\xi(34-39\xi)}{27\ep_\gamma}-\frac{2}{81}(695-798\xi+891\xi^2)
\nonum \\
& +\frac{\pi^2}{54}(71-21\xi(2-3\xi))+\Ord(\ep_\gamma)\Bigg]\, .
\end{flalign}
\end{subequations}
}

From \eqref{Sigma2:S:a+b+c:exp} and \eqref{Sigma2:V:a+b+c:exp} together with (\ref{def:Zpsi+Zm}), 
the two-loop contributions to the renormalization constants read:
\begin{subequations}
\label{Z2:psi+m}
\begin{flalign}
\delta Z_{2m} & = 
\frac{2(3-2\ep_e)^2}{(1-\ep_e)^2(2-\ep_e)^2\,\ep_\gamma^2}
+\frac{2}{\ep_\gamma}\,\left(\frac{2N K_1}{2-\ep_e}-\frac{3-2\ep_e}{(1-\ep_e)^2 (2-\ep_e)^3}\right) \, , 
\label{Z2:m} \\
\delta Z_{2\psi} & = 
\frac{(\ep_e-(2-\ep_e)\,\xi_r)^2}{2(2-\ep_e)^2\,\ep_\gamma^2}
 - \frac{2}{\ep_\gamma}\,\left(\frac{N \ep_e K_1}{2-\ep_e} + \frac{(3-2\ep_e)(\ep_e(3-\ep_e)-1)}{(1-\ep_e)(2-\ep_e)^3}\right) \, . 
\label{Z2:psi}
\end{flalign}
\end{subequations}
From \eqref{gen:gammaX}, the corresponding two-loop contribution to the anomalous dimensions read:
\begin{subequations}
\label{gamma2:psi+m}
\begin{flalign}
\gamma_{2m} & = -8\,\left(\frac{2N\,K_1}{2-\ep_e}-\frac{3-2\ep_e}{(1-\ep_e)^2 (2-\ep_e)^3}\right) \, , 
\label{gamma2:m} \\
\gamma_{2\psi} & = 8\,\left(\frac{N\,\ep_e K_1}{2-\ep_e} + \frac{(3-2\ep_e)(\ep_e(3-\ep_e)-1)}{(1-\ep_e)(2-\ep_e)^3}\right) \, . 
\label{gamma2:psi}
\end{flalign}
\end{subequations}

\subsection{Anomalous dimensions}
\label{sec:AD}

Combining the above derived one-loop \eqref{gamma1:psi+m} and two-loop \eqref{gamma2:psi+m} contributions to the anomalous dimensions yields:
\begin{subequations}
\label{gamma:psi+m}
\begin{flalign}
\gamma_{m} & = 4\bar{\al}_r\,\frac{3-2\ep_e}{(2-\ep_e)(1-\ep_e)} - 8\bar{\al}_r^2\,\left(\frac{2N\,K_1}{2-\ep_e}-\frac{3-2\ep_e}{(1-\ep_e)^2 (2-\ep_e)^3}\right) 
+ \Ord(\bar{\al}_r^3) \, , 
\label{gamma:m} \\
\gamma_{\psi} & = 2\bar{\al}_r\,\left( \xi_r - \frac{\ep_e}{2-\ep_e} \right) + 8\bar{\al}_r^2\,\left(\frac{N\,\ep_e K_1}{2-\ep_e} + \frac{(3-2\ep_e)(\ep_e(3-\ep_e)-1)}{(1-\ep_e)(2-\ep_e)^3}\right) + \Ord(\bar{\al}_r^3)\, , 
\label{gamma:psi}
\end{flalign}
\end{subequations}
where $\gamma_m$ is fully gauge invariant as it should be while the gauge-variance of $\gamma_\psi$ is at one-loop in accordance
with the Landau-Khalatnikov-Fradkin transformation, see, \eg, \cite{James:2019ctc}. Interestingly, \eqref{gamma:m} and \eqref{gamma:psi} have similar structures.
In particular, we see that in the limit of RQED$_{4,3}$ ($\ep_e \ra 1/2$) the factors of $\pi^2$ arise from $K_1$, see \eqref{params-RQED4de-Lp+K1},  and not
from the $\ep_\gamma$-expansion of $\Gamma$-functions. It turns out however that, because of these $K_1$ terms, \eqref{gamma:psi+m}
(and similarly for the two-loop renormalization constants \eqref{Z2:psi+m} and self-energies 
\eqref{Sigma2:S:a+b+c:exp} and \eqref{Sigma2:V:a+b+c:exp} above) do not apply to the case of QED$_4$. 
As discussed below \eqref{Pi1:exp}, this discrepancy (which is even more severe for $\gamma_m$ than $\gamma_\psi$)  
originates from the fact that the limits $\ep_e \ra 0 $ and $\ep_\gamma \ra 0$ do not
commute for the one-loop polarization operator and hence for \eqref{Sigma2Sc:exp} and \eqref{Sigma2Vc:exp} where it appears as a subgraph.\footnote{This problem will of course also appear at higher orders for each diagram containing a fermion loop.}
 So \eqref{gamma:psi+m} only apply to RQED$_{4,3}$.

The expressions \eqref{gamma:psi+m} can be improved in order to cover both cases of QED$_4$ and RQED$_{4,3}$ 
at the expense of introducing an additional parameter $z$ such that
\be
\text{L}(\Sigma, d_e,z) =z\, \text{L}(\Sigma,d_e=4) + (1-z)\, \text{L}(\Sigma,d_e<4), \qquad z=\delta_{\ep_e,0} \, ,
\label{exp:improved}
\ee
where $\text{L}(\Sigma,d)$ stands for Laurent expansion of $\Sigma = \{\Pi_1,\Sigma_{2Va},\Sigma_{2Sa}\}$ near the dimension $d$. 
All calculations done, the two-loop contributions to the renormalization constants \eqref{Z2:psi+m} now take the 
form:
\begin{subequations}
\label{Z2:psi+m:improved}
\begin{flalign}
\delta Z_{2m} & = 
\frac{2}{\ep_\gamma^2}\,\left(\frac{(3-2\ep_e)^2}{(1-\ep_e)^2(2-\ep_e)^2} - z N \right)
+\frac{2}{\ep_\gamma}\,\left(\frac{2(1-z)N K_1}{2-\ep_e}+ \frac{5 z N}{6}-\frac{3-2\ep_e}{(1-\ep_e)^2 (2-\ep_e)^3}\right) \, , 
\label{Z2:m:improved} \\
\delta Z_{2\psi} & = 
\frac{(\ep_e-(2-\ep_e)\,\xi_r)^2}{2(2-\ep_e)^2\,\ep_\gamma^2}
 - \frac{2}{\ep_\gamma}\,\left(\frac{(1-z) N \ep_e K_1}{2-\ep_e} - \frac{z N}{2} + \frac{(3-2\ep_e)(\ep_e(3-\ep_e)-1)}{(1-\ep_e)(2-\ep_e)^3}\right) \, , 
\label{Z2:psi:improved}
\end{flalign}
\end{subequations}
where, in QED$_4$ the running of the coupling constant at one-loop has been taken into account via: 
$\gamma_A = 8 z N \bar{\al}_r /3 + \Ord(\bar{\al}_r^2)$. Hence, we obtain one of the central results of this paper in the form of improved
expressions for the anomalous dimensions in RQED$_{4,d_e}$:
\begin{subequations}
\label{gamma:psi+m:improved}
\begin{flalign}
\gamma_m & = 4\bar{\al}_r\,\frac{3-2\ep_e}{(2-\ep_e)(1-\ep_e)}
- 8 \bar{\al}_r^2\,\Bigg( \frac{2 (1-z) N K_1}{2-\ep_e} + \frac{5 z N}{6} - \frac{3-2\ep_e}{(2-\ep_e)^3(1-\ep_e)^2} \Bigg)+\Ord(\bar{\al}_r^3)\, ,
\label{gamma:m:improved} \\
\gamma_\psi
& = 2\bar{\al}_r\,\left(\xi_r - \frac{\ep_e}{2-\ep_e}\right)
 +8\bar{\al}_r^2\,\Bigg(\frac{(1-z) N \ep_e K_1}{2-\ep_e} - \frac{z N}{2} +    \frac{(3-2\ep_e)(\ep_e(3-\ep_e)-1)}{(2-\ep_e)^3(1-\ep_e)} \Bigg)
+\Ord(\bar{\al}_r^3)\, .
\label{gamma:psi:improved}
\end{flalign}
\end{subequations}

A remark is in order here in relation with \eqref{gamma:psi:improved}. Within the SD formalism, 
see, \eg, the review \cite{Roberts:1994dr}, a common procedure to simplify the equations
and minimize the gauge-variance of the solutions is to consider the gauge for which the fermion anomalous
dimension vanishes. Such a gauge is referred to as the ``good gauge''. At one-loop, from \eqref{gamma:psi:improved}, the ``good gauge'' is simply:
\be
\xi_g = \frac{\ep_e}{2-\ep_e}\, ,
\label{xi:good:one-loop}
\ee
which corresponds to the Landau gauge ($\xi_g=0$) in QED$_4$ and the Nash gauge ($\xi_g=1/3$, see \cite{Nash:1989xx}) in RQED$_{4,3}$.
Proceeding in a similar way at two-loop, with the help of \eqref{gamma:psi:improved}, we find that the ``good gauge'' becomes:
\be
\xi_g=\frac{\ep_e}{2-\ep_e}+\bar{\al}_r\,\left(\frac{4(3-2\ep_e)(1-3\ep_e+\ep_e^2)}{(2-\ep_e)^3(1-\ep_e)}
-\frac{4N(1-z)\ep_e K_1(\ep_e)}{2-\ep_e}+2Nz\right)+\Ord(\bar{\al}_r^3)\, .
\label{xi:good:two-loop}
\ee
So, in the case of RQED$_{4,3}$ ($z = 0$ and $\ep_e \ra 1/2$), \eqref{xi:good:two-loop} yields:
\be
\xi_g = \frac{1}{3} - 8\bar{\al}_r\,\left( N \zeta_2 + \frac{4}{27} \right)\, ,
\label{xi:good:two-loop:RQED43}
\ee
and in the case of QED$_{4}$ ($z = 1$ and $\ep_e \ra 0$), the (two-loop) ``good gauge'' reads:
\be
\xi_g = \bar{\al}_r\,\left( 2 N + \frac{3}{2} \right)\, .
\label{xi:good:two-loop:QED3}
\ee
Note that, in the case of QED$_4$, \eqref{xi:good:two-loop:QED3} is known since a long time (see, \eg, eq.~(B.1) in \cite{Johnson:1964da} for an early derivation in the quenched case) and can presently be extended to 5 loops thanks to state of the art results for $\gamma_\psi$ \cite{Luthe:2016xec,Baikov:2017ujl}.

We may now proceed in applying \eqref{Z2:psi+m:improved} and \eqref{gamma:psi+m:improved} to various cases of interest. Let's consider first the most
important case of RQED$_{4,3}$ which amounts to set $z = 0$ and $\ep_e \ra 1/2$. From \eqref{Z1:psi+m} and \eqref{Z2:psi:improved}, 
the renormalization constants read:
\begin{subequations}
\label{Z:psi+m:improved:RQED43}
\begin{flalign}
Z_m & =1 - \frac{16\, \bar{\al}_r}{3 \, \ep_\gamma} +  
\bar{\al}_r^2\,\left( \frac{128}{9 \, \ep_\gamma^2} + \frac{16}{\ep_\gamma}\,\left( N \zeta_2 - \frac{8}{27} \right) \right) + \Ord(\bar{\al}_r^3)\, ,
\label{Z:m:improved:RQED43} \\
Z_\psi & = 1 + \bar{\al}_r\,\frac{1 - 3 \xi_r}{3 \, \ep_\gamma} 
+ \bar{\al}_r^2\,\left( \frac{(1 - 3 \xi_r)^2}{18 \, \ep_\gamma^2} - \frac{4}{\ep_\gamma}\,\left( N \zeta_2 + \frac{4}{27} \right) \right) + \Ord(\bar{\al}_r^3)\, .
\label{Z:psi:improved:RQED43}
\end{flalign}
\end{subequations}
And from \eqref{gamma:psi+m:improved}, the anomalous dimensions read:
\begin{subequations}
\label{gamma:psi+m:improved:RQED43}
\begin{flalign}
\gamma_m & =\frac{32\, \bar{\al}_r}{3} - 
64\,\bar{\al}_r^2\,\left( N \zeta_2 - \frac{8}{27} \right) + \Ord(\bar{\al}_r^3)\, ,
\label{gamma:m:improved:RQED43} \\
\gamma_\psi & = -2 \,\bar{\al}_r\,\frac{1 - 3 \xi_r}{3} 
+16\,\bar{\al}_r^2\,\left( N \zeta_2 + \frac{4}{27} \right) + \Ord(\bar{\al}_r^3)\, ,
\label{gamma:psi:improved:RQED43}
\end{flalign}
\end{subequations}
where \eqref{gamma:psi:improved:RQED43} agrees with the result of \cite{Kotikov:2013eha}. Our result \eqref{gamma:m:improved:RQED43} is new and corresponds to the parity-even mass anomalous dimension of RQED$_{4,3}$ for an arbitrary number $N$ of 4-component spinors.\footnote{Let's note that, in \cite{DiPietro:2019hqe}, the parity-odd mass anomalous dimension of a single 2-component spinor was computed. The computation of such a quantity requires taking into account of the fact that the trace of three $2\times2$ Dirac gamma matrices is non-zero: $\Tr[\gamma^\mu \gamma^\nu \gamma^\rho] = 2\I\, 
\ep^{\mu \nu \rho}$ where $\ep^{\mu \nu \rho}$ is the rank 3 totally antisymmetric tensor. This affects the one-loop polarization operator and hence the coefficient of $\zeta_2$ in $\gamma_m$ through the scalar part of diagram (a) of fig.~\ref{fig:fig2} (and corresponding self-energy $\Sigma_{2Sa}(p^2)$). Recomputing this diagram with an arbitrary number, $n$, of 2-component spinors (which is such that $n=2N$) yields:
\be
\gamma_m = \frac{32\,\bar{\al}_r}{3} 
- 64\,\bar{\al}_r^2\,\left( n \,\frac{1+3\mathcal{T}_n}{2}\,\zeta_2 - \frac{8}{27} \right) + \Ord(\bar{\al}_r^3)\, ,
\label{gamma:m:improved:RQED43:all}
\ee
where $\mathcal{T}_n$ arises from the epsilon tensors and is such that $\mathcal{T}_{2k}=0$ while 
$\mathcal{T}_{2k+1}=1$ for any integer $k$. Eq.~\eqref{gamma:m:improved:RQED43:all} is a general formula including both cases of an odd or an even number of 2-component spinors. Indeed, for $n=2N$, we recover our eq.~\eqref{gamma:m:improved:RQED43} while
for $n=1$ we recover the result of \cite{DiPietro:2019hqe} which, in our notations, reads: 
$\gamma_m = 32\,\bar{\al}_r/3 
- 128\,\bar{\al}_r^2\,( \zeta_2 - 4/27) + \Ord(\bar{\al}_r^3)$.
Note also that the model considered in \cite{DiPietro:2019hqe} is that of a reflective boundary (say, with fine structure constant $\al_{\text{bdry}}$) while our case corresponds to a transparent interface 
(with fine structure constant $\al$); these models are related by: $\al_{\text{bdry}} = \al/2$.
}
In order to get more confidence in our result for $\gamma_m$, we proceed on showing that it can be mapped to the one of QED$_3$ at 
NLO in the large-$N$ expansion with the help of \cite{Kotikov:2016yrn}: 
\be
\label{mappingQED43toQED3}
\bar{\al}_r\rightarrow \frac{1}{N\,\pi^2}, \qquad 
\hat{\Pi}_1^{(4,3)} = 2 N\pi^2 \rightarrow \hat{\Pi}_2 = 4\,\left(\frac{92}{9}-\pi^2 \right) \, ,
\ee
where $\hat{\Pi}_1^{(4,3)}$ was defined in \eqref{Pi1:exp:RQED43} and $\hat{\Pi}_2$ computed in 
\cite{Gusynin:2000zb,Teber:2012de,Kotikov:2013kcl}.\footnote{Notice that this result was also known to the authors of
\cite{Kotikov:1989nm,Kotikov:2011kg,Gracey:1993iu}. Though it did not appear explicitly in these papers, 
the knowledge of $\hat{\Pi}_2$ was required to perform the calculations carried out in these papers.}
Substituting \eqref{mappingQED43toQED3} in our RQED$_{4,3}$ result \eqref{gamma:m:improved:RQED43} yields:
\be
\gamma_m = \frac{32}{3\pi^2N} + \frac{128}{\pi^4 N^2}\,\left( \zeta_2 - \frac{14}{9} \right) +\Ord ( 1/N^3 )\, ,
\label{gammam:QED3:NLO}
\ee
which corresponds exactly to the result obtained by Gracey in \cite{Gracey:1993sn}.\footnote{In addition, the mapping of
\cite{Kotikov:2016yrn} allows to recover \eqref{gamma:psi:improved:RQED43} from the field anomalous dimension of QED$_3$ at
NLO in the large-$N$ expansion which can also be extracted from Gracey's work \cite{Gracey:1993sn} and that we provide here for completeness:
\be
\gamma_\psi = \frac{4\,(2-3\xi)}{3\,\pi^2 N} - \frac{8}{3\,\pi^4 N^2}\,\left( \frac{8}{9} + \frac{2-3\xi}{4}\,\hat{\Pi}_2 \right) + \Ord ( 1/N^3 )\, ,
\label{gammapsi:QED3:NLO}
\ee
where $\hat{\Pi}_2$ was defined in \eqref{mappingQED43toQED3}.
Hence, all results are in complete agreement. Note that from 
\eqref{gammapsi:QED3:NLO} we may also compute the``good gauge'' in large-$N$ QED$_3$ at NLO that reads:
\be
\xi_g = \frac{2}{3}\,\left(1 - \frac{8}{9\,\pi^2 N} \right)\, ,
\ee
in agreement with eq.~(4.8) in \cite{Gusynin:2016som}.
}


Our results also allow us to consider the well-known case of QED$_{4}$, see, \eg, the textbook \cite{Grozin:2005yg}, thereby strengthening their validity. 
From \eqref{Z1:psi+m} and \eqref{Z2:psi:improved}, this amounts to set $z = 1$ and $\ep_e \ra 0$ yielding:
\begin{subequations}
\label{Z:psi+m:improved:QED4}
\begin{flalign}
Z_m & =1 - \frac{3 \bar{\al}_r}{\ep_\gamma} + \frac{\bar{\al}_r^2}{2}\,\left( \frac{9 - 4N}{\ep_\gamma^2} + 
\frac{1}{2\,\ep_\gamma}\,\left(\frac{20 N}{3} - 3 \right) \right) + \Ord(\bar{\al}_r^3)\, ,
\label{Z:m:improved:QED4}
\\
Z_\psi & = 1 - \frac{\bar{\al}_r\,\xi_r}{\ep_\gamma} + \frac{\bar{\al}_r^2}{2}\,\left(  \frac{\xi_r^2}{\ep_\gamma^2} + 
\frac{4N + 3}{2\,\ep_\gamma} \right) + \Ord(\bar{\al}_r^3)\, .
\label{Z:psi:improved:QED4}
\end{flalign}
\end{subequations}
Proceeding in a similar way from \eqref{gamma:psi+m:improved}, the corresponding anomalous dimensions read: 
\begin{subequations}
\label{gamma:psi+m:improved:QED4}
\begin{flalign}
\gamma_m & =6\,\bar{\al}_r - \bar{\al}_r^2\left(\frac{20 N}{3} - 3 \right) + \Ord(\bar{\al}_r^3)\, ,
\label{gamma:m:improved:QED4}
\\
\gamma_\psi & =2\,\bar{\al}_r\,\xi_r - \bar{\al}_r^2\,\left(4 N + 3 \right) + \Ord(\bar{\al}_r^3)\, .
\label{gamma:psi:improved:QED4}
\end{flalign}
\end{subequations}
In \eqref{gamma:m:improved:QED4}, the one-loop contribution was probably first computed in \cite{Johnson:1964da}, the quenched two-loop one in \cite{Baker:1971yc} and the unquenched two-loop contribution can be extracted from the QCD calculations of \cite{Tarrach:1980up,Nachtmann:1981zg} as we could learn from 
\cite{Tarasov:2019rwk} where the three-loop QCD calculation was performed. Notice that, presently, $\gamma_m$ is known up to five
loops in QCD \cite{Luthe:2016xec,Baikov:2017ujl} from which the corresponding QED result can be extracted (we shall use it in the next section).\footnote{Let's note 
that \eqref{gamma:psi+m:improved} seems to also apply to the case of QED$_{4,1}$ ($z = 0$ and $\ep_e\rightarrow3/2$) 
for which $\gamma_m  = 0 + \Ord(\bar{\al}_r^3)$ and $\gamma_\psi = 2 \bar{\al}_r\,(\xi_r - 3) + \Ord(\bar{\al}_r^3)$.
%
%
 This limit corresponds to the quantum mechanics of a point particle coupled to a fully retarded four-dimensional electromagnetic field.
}

\section{Critical properties}
\label{sec.DMG}

As explained in the Introduction, a standard approach to study the critical properties of gauge theories is by solving the SD equations.
Truncating these equations at LO is generally unsatisfactory as the resulting critical coupling may be strongly gauge-variant. The fully gauge-invariant
procedure advocated in the recent \cite{Gusynin:2016som} and \cite{Kotikov:2016prf} for large-$N$ QED$_3$ requires computing NLO corrections and then performing a so-called Nash
resummation (following the seminal work of Nash \cite{Nash:1989xx}) in order to properly cancel the gauge dependence both at LO and NLO. The simplicity
of the resulting gap equation (from which the gauge-invariant critical coupling is extracted) drastically contrasts with the complexity
of the calculations that need to be performed in order to derive it. Following \cite{Gusynin:2016som}, in this section we 
provide a semi-phenomenological construction of the gap equation for 
RQED$_{4,d_e}$ and apply it to cases of interest. Before that, we would like nevertheless to apply the SD formalism at LO for RQED$_{4,d_e}$. This will allow
us to illustrate some of the difficulties we have just mentioned but also to underline 
the importance of the ``good gauge'' introduced in the previous section.

\subsection{Leading order solution of SD equations for RQED$_{4,d_e}$}
\label{sec.DMG:SD:LO}

The SD equation for the fermion self-energy in RQED$_{4,d_e}$ reads:
\be
-\I\, \Sigma(p) = \mu^{2\ep_\gamma} \! \! \int [\D^{d_{e}} k]\, \Gamma^\mu\,S(k)\,\Gamma_0^\nu\,D_{\mu \nu}(p-k)\, ,
\label{Sigma:SDeq}
\ee
where $S(k)$, $\Gamma^\mu$ and $D_{\mu \nu}(k)$ are the full fermion propagator, vertex function and  photon propagator, respectively. In general, \eqref{Sigma:SDeq} is
coupled to the Dyson equation for the vertex function and the photon propagator. A dynamical
mass arises as a non-trivial solution of this system of coupled equations for 
$\Sigma_S(p^2)$ that needs to be determined self-consistently. 

Focusing on the LO approximation in the coupling constant, all equations decouple as both the vertex function and the photon propagators are taken as the free ones: $\Gamma^\mu=\Gamma_0^\mu$ and $D^{\mu\nu}(p)=D_0^{\mu\nu}(p)$. Moreover, the wave-function renormalization is neglected, \ie, $\Sigma_V=0$, which implies that $S(p)=\I(\Sp - \Sigma_S)^{-1}$ where $\Sigma_S$ is now the dynamically 
generated parity-conserving mass (for convenience, the mass parameter has been absorbed in $\Sigma_S$). With the  help of the parametrization \eqref{def:SigmaV+SigmaS} (with $m=1$),  \eqref{Sigma:SDeq} significantly simplifies and can be written as:
\be
\Sigma_S(p^2) = \frac{-\I \, e^2 \Gamma(1-\ep_e)}{(4\pi)^{\ep_e}}\,(d_e-1+\tilde{\xi})\, \mu^{2\ep_\gamma}\!\! 
\int \frac{[\D^{d_{e}} k]\, \Sigma_S(k^2)}{[-k^2 + \Sigma_S^2(k^2)]\,[-(p-k)^2]^{1-\ep_e}}\, .
\label{Sigma:selfconsistentSD}
\ee
At the critical point, \eqref{Sigma:selfconsistentSD} can be linearized by taking the limit 
$\Sigma_S^2(k^2) \ll k^2$ and a power-law ansatz can be taken for the mass function $\Sigma_S(p^2)$ 
such that
\be
\Sigma_S(p^2) = m_{\text{dyn}}\,(-p^2)^{-a/2}\, ,
\label{SigmaS:ansatz}
\ee
where the mass is assumed to be dynamical in origin and the index $a$ has to be self-consistently determined. 
As first noticed by Kotikov \cite{Kotikov:1989nm,Kotikov:2011kg}, together with \eqref{SigmaS:ansatz}, the linearized eq.~\eqref{Sigma:selfconsistentSD} is now a massless integral which is very easily 
solved in dimensional regularization with the help of \eqref{App:1-loop:J}. The
solution being finite for all $d_e$, one can set $\ep_\gamma \ra 0$ and straightforwardly 
derive the LO gap equation:
\be
a \big( d_e - 2 - a \big)= 4 \bar{\alpha}\,\big(3-\ep_e + (1-\ep_e)\xi \big) 
+ \Ord(\bar{\alpha}^2)\, .
\label{gap-equation:LO}
\ee
Solving it yields two values for the index $a$:
\be
a_\pm = (1-\ep_e)\,\left(1 \pm \sqrt{1 - 
4\bar{\alpha}\, \frac{3-\ep_e + (1-\ep_e)\xi}{(1-\ep_e)^2}}\right) \, .
\ee
Dynamical mass generation takes place for complex values of $a$, \ie, for $\alpha>\alpha_c$ with
\be
\alpha_c(\xi)=\frac{\pi(1-\ep_e)^2}{3-\ep_e+(1-\ep_e)\xi}\, .
\label{alphac:LO:xi}
\ee
Notice that, in the case of RQED$_{4,3}$ ($\ep_e=1/2$), \eqref{alphac:LO:xi} leads
to: $\alpha_c=\pi/(2(5+\xi))$ in agreement with \cite{Kotikov:2016yrn}. On the other hand, 
in the case of QED$_4$ ($\ep_e=0$), we obtain $\alpha_c=\pi/(3+\xi)$, which exactly corresponds to the result of Rembiesa \cite{Rembiesa:1990jd} according to \cite{Atkinson:1993mz}.

As anticipated, \eqref{alphac:LO:xi} has a strong gauge dependence which is not
satisfactory. In order to minimize it, we consider the ``good gauge'' which is given by \eqref{xi:good:one-loop} at one-loop. This yields:
\be
\al_{c}(\xi_g) = \frac{\pi\,(2-\ep_e)(1-\ep_e)^2}{2\,(3-2\ep_e)}\, .
\label{alc:fromSD}
\ee
We shall come back to this result and discuss it in detail in the next subsections. 
At this point, let's note that the LO gap equation itself, \eqref{gap-equation:LO}, 
can also be written in the ``good gauge'' where it may be expressed in the form:
\be
a\,(d_e-2 - a) = (d_e - 2)\,\big( \gamma_{1m} \bar{\alpha} + \Ord(\bar{\alpha}^2) \big)\, .
\label{gap-equation:LO:goodxi}
\ee
Interestingly, the right hand side of \eqref{gap-equation:LO:goodxi} involves the one-loop mass anomalous dimension \eqref{gamma1:psi+m} --- a gauge-invariant quantity --- 
as the only input. 

The powerful technique of Kotikov \cite{Kotikov:1989nm,Kotikov:2011kg},  that we have used here to easily solve the LO SD equation 
in dimensional regularization, can possibly be extended to higher orders for QED$_4$ along the lines
of the recent \cite{Kotikov:2016prf}. Of course, an NLO computation is more complicated
and out of the scope of the present paper. Instead, in the following subsection, we
will present general arguments allowing us to build a fully gauge-invariant gap equation
that is valid at any order in the coupling constant thereby generalizing 
\eqref{gap-equation:LO:goodxi}.

\subsection{Gap equation and criterion for dynamical mass generation}

We start by recalling that, from the operator product expansion, the scalar part of the fermion self-energy has two asymptotes in the UV \cite{Lane:1974he,Politzer:1976tv,Miransky:1985wzx,Cohen:1988sq}
(see also the textbook \cite{Miransky:1994vk} sec.~12.3):
\be
\Sigma_S(p) \sim m\,p_E^{-\gamma_m} + m_{\text{dyn}}\,p_E^{-(d_e-2-\gamma_m)}\, ,
\label{Sigma:asymptotics}
\ee
where $p_E^2=-p^2$ is the Euclidean momentum, $m$ is the bare mass and $m_{\text{dyn}}$ the dynamical one. As noticed in \cite{Cohen:1988sq},
dynamical mass generation arises from the coalescence of these two asymptotes. In particular, deep in the UV, $p_E^2 \ra \infty$, 
the dynamical mass is favoured over the bare mass provided the mass anomalous dimension is large enough: $\gamma_m > (d_e-2)/2$.
The non-perturbative criterion for dynamical mass generation is therefore given by $\gamma_m(\al_{c}) = (d_e-2)/2$ which requires the knowledge
of the exact $\gamma_m$. 

In our case, we only have access to a perturbative expansion for $\gamma_m$. Hence, we would like to find a criterion 
for dynamical mass generation (which is intrinsically a non-perturbative mechanism) that could be truncated at a given order of the perturbative expansion of $\gamma_m$.
This can be achieved with the help of the gap equation found from the SD formalism. Actually, as we saw in the previous subsection, 
we know that the all order ansatz (valid at the critical point) for the fermion self-mass 
is given by $\Sigma_S(p) \sim p_E^{-a}$ (see \eqref{SigmaS:ansatz}) where the index $a$ is determined self-consistently. 
Comparing this ansatz to \eqref{Sigma:asymptotics}, and in particular to the second asymptote of this equation, we see that $a = d_e - 2 -\gamma_m$ and is therefore
related to the mass anomalous dimension. From \cite{Gusynin:2016som}, we learn that the gap equation (which is quadratic in $a$) is built
up from the two asymptotes of \eqref{Sigma:asymptotics} (at least for large-$N$ QED$_3$ at NLO in the $1/N$-expansion). 
Extending the result of Gusynin and Pyatkovskiy \cite{Gusynin:2016som} to arbitrary $d_e$,
we therefore assume that, for RQED$_{4,d_e}$, the gap equation 
is quadratic in $a$ at all loop orders and takes the form: 
\be
a(d_e-2-a) = \gamma_m(d_e -2 - \gamma_m)\, ,
\label{gap-equation}
\ee
with $\gamma_m$ as the only input. As we saw in the previous subsection, in the SD formalism, the dynamical mass is generated when $a$ becomes complex, \ie, for $(a - (d_e-2)/2)^2<0$. 
In terms of the mass anomalous dimension and at the critical point, this translates into:
\be
\left(\gamma_m(\al_{c})-\frac{d_e-2}{2}\right)^2=0\, ,
\label{gap-eqn:de}
\ee
from which the critical coupling $\al_c$ can be computed. Since $\gamma_m$ is gauge invariant, the resulting critical coupling 
will automatically be gauge invariant too. And as it is built from the SD formalism, it can be truncated to the accuracy 
at which $\gamma_m$ is known (Eq.~\eqref{gap-equation} reduces to \eqref{gap-equation:LO:goodxi} at the
LO in $\al$). From this polynomial equation, we will obtain multiple solutions for $\al_c$. In the following, 
the physical $\al_c$ will be taken as the the smallest positive real solution that is found. 

At this point, we would like to comment on the fact that the large mass anomalous dimension required for dynamical mass generation also affects 
the dimension of the quartic fermion operator $\Delta[(\bar{\psi} \psi)^2] = 2d_e - 2 - \gamma_{(\bar{\psi} \psi)^2}$ where
$\gamma_{(\bar{\psi} \psi)^2}$ is the associated anomalous dimension. Indeed, assuming that $\gamma_{(\bar{\psi} \psi)^2} = 2\gamma_m$,
we have $\Delta[(\bar{\psi} \psi)^2] = 2d_e - 2 - 2\gamma_m \leq d_e$ for $\gamma_m \geq (d_e-2)/2$, \ie, the quartic fermion operator is marginal
at the critical point and becomes relevant once the dynamical mass is generated. However, from these arguments, 
the marginality of the quartic fermion operator at criticality appears as a consequence of \eqref{gap-eqn:de} and holds only in an approximate way.
According to the literature on QED$_4$, see, \eg, the review \cite{Roberts:1994dr} as well as \cite{Bardeen:1985sm,Leung:1985sn,Miransky:1988gk,Leung:1989hw}, 
the assumption $\gamma_{(\bar{\psi} \psi)^2} = 2\gamma_m$ is supposed to be valid 
in the quenched and rainbow approximation.\footnote{In the rainbow (or ladder) approximation, the full vertex $\Gamma^\mu$ entering the SD equations
is taken as the free vertex, $\Gamma^\mu = -\I e \gamma^\mu$. Note that our (two-loop) perturbative calculations do take into account of vertex corrections and 
our analysis is therefore beyond the rainbow approximation.} In the case of QED$_4$, there is evidence
that it still holds beyond the rainbow approximation \cite{Atkinson:1993mz}. Note also that the quenched approximation significantly affects QED$_4$ because
in this approximation the coupling does not run. But its effect is weaker for a ``standing'' theory such as 
RQED$_{4,3}$. These arguments suggest that the marginality of the quartic fermion operator at criticality may hold even beyond 
the quenched (especially for RQED$_{4,3}$) and rainbow approximations. A more careful study of the validity of this approximation 
goes beyond the scope of this paper.

In the next sections, we will apply \eqref{gap-eqn:de} to the study of the critical properties of RQED$_{4,d_e}$ with applications to
RQED$_{4,3}$ and QED$_{4}$. For the case of RQED$_{4,3}$, we will find perfect agreement with the SD formalism \cite{Kotikov:2016yrn}
which is natural since \eqref{gap-eqn:de} is built from the SD formalism for large-$N$ QED$_3$ \cite{Gusynin:2016som} 
which can be mapped to RQED$_{4,3}$ \cite{Kotikov:2016yrn}. As for QED$_4$, our approach is more phenomenological 
since it amounts to extrapolate an equation valid for RQED$_{4,3}$ to the more subtle case of a ``running'' theory. 

\subsection{Application to quenched RQED$_{4,d_e}$}

We first consider RQED$_{4,d_e}$ at one-loop. Substituting the one-loop expression of $\gamma_m$, \eqref{gamma1:psi+m}, in \eqref{gap-eqn:de}
yields the following critical coupling constant:
\be
\al_{c} = \frac{\pi\,(2-\ep_e)(1-\ep_e)^2}{2\,(3-2\ep_e)}\, ,
\label{alc:RQED4de:one-loop}
\ee
which corresponds exactly to the result \eqref{alc:fromSD} obtained via the SD approach in the ``good gauge'' \eqref{xi:good:one-loop}.
In the case of RQED$_{4,3}$ ($\ep_e = 1/2$), \eqref{alc:RQED4de:one-loop} yields:
$\al_{c} = 3\pi/32 =0.2945$, 
%
%
 which agrees with the result of \cite{Gorbar:2001qt,Kotikov:2016yrn}. 
Note that from the mapping (\ref{mappingQED43toQED3})
to large-$N$ QED$_3$, the corresponding critical fermion flavour number reads: $N_{c}=128/(3\pi^2)=4.3230$, 
in agreement with \cite{Nash:1989xx,Gorbar:2001qt,Gusynin:2016som,Kotikov:2016prf}. 

In the case of QED$_{4}$ ($\ep_e = 0$), we recover the celebrated result: 
$\al_{c}= \pi/3=1.0472$, which was obtained via the rainbow approximation in the Landau gauge (which is the ``good gauge'' for QED$_4$ at one loop) 
in the early papers \cite{Fukuda:1976zb,Fomin:1984tv,Miransky:1984ef}. 
 In \cite{Atkinson:1993mz}, a numerical solution of SD equations with Curtis-Pennington vertex (and hard cut-off regularization) led to $\al_{c}(\xi) \approx 0.93$ with variations of the critical coupling of only a couple of percent over a wide range of the gauge fixing parameter; such a result deviates by $11\%$ from $\pi/3$.
In \cite{Schreiber:1998ht}, calculations using dimensional regularization, which is a gauge-invariant
regularization scheme, were  found to agree with the hard cut-off regularization ones to within numerical precision. Moreover,
 in \cite{Gusynin:1998se} the result $\al_{c}= \pi/3$ was extracted analytically from dimensionally regularized SD equations
in the Landau gauge in perfect agreement with our own calculation of sec.~\ref{sec.DMG:SD:LO}.
Based on \eqref{gap-eqn:de}, we obtained in a very simple way that $\al_{c}= \pi/3$ is actually the fully gauge-invariant result at LO. This motivates us 
to include higher order corrections which is quite straightforward in our formalism provided $\gamma_m$ is known. 
%
%

We therefore consider RQED$_{4,d_e}$ at two-loop. Substituting the two-loop expression of $\gamma_m$, \eqref{gamma:m:improved}, in \eqref{gap-eqn:de}, and selecting as a physical critical coupling the smallest
value obtained, yields the following result:
\be
\al_{c}(N) = \frac{\pi\,(2-\ep_e)(1-\ep_e)^2}{3 - 2\ep_e + \sqrt{\Delta_{N}}}\, , 
\label{alc:RQED4de:two-loop}
\ee
where
\be
\Delta_{N}=\frac{(1-\ep_e)(3-2\ep_e)}{2-\ep_e} - 2N(1-z)(2-\ep_e)(1-\ep_e)^3 K_1 - \frac{10 N z}{3}\, .
\label{DeltaN:def}
\ee
It turns out that $\al_{c}$ in \eqref{alc:RQED4de:two-loop} is actually complex and hence unphysical for all integer values of $N$ except $N=0$. 
This comes from the fact that $\Delta_N$ in \eqref{DeltaN:def} is positive only for values of $N$
which are smaller than:
\be
N_{\text{max}} = \frac{3\,(3-2\ep_e)}{2\,\big( 10z + {3}(1-z)(1-\ep_e)^2(2-\ep_e)^2\,K_1 \big)}\, ,
\label{Nmax}
\ee
and this maximal value is smaller than $1$ for $d_e \leq4$. Thus, the critical coupling \eqref{alc:RQED4de:two-loop} is defined only in the case of quenched RQED$_{4,d_e}$ where its expression simplifies.
We display it explicitly for clarity:
\be
\al_{c}(N=0) = \frac{\pi\,(2-\ep_e)(1-\ep_e)^2}{3 - 2\ep_e + \sqrt{\Delta_{0}}}\, , \quad \Delta_{0}=\frac{(1-\ep_e)(3-2\ep_e)}{2-\ep_e}\, ,
\label{alc:RQED4de:two-loop:quenched}
\ee
where, together with $N$, all $z$-dependence dropped out. In the case of RQED$_{4,3}$ ($\ep_e = 1/2$), \eqref{alc:RQED4de:two-loop:quenched} yields: 
$\al_{c}(N=0)=9\pi/(8\,(6+\sqrt{6}))=0.4183$, in agreement with \cite{Kotikov:2016yrn}. Note that from the mapping (\ref{mappingQED43toQED3})
to large-$N$ QED$_3$, the corresponding critical fermion flavour number reads: $N_{c}=16(4+\sqrt{3\pi^2-28})/(3\pi^2)=2.8470$, in agreement with \cite{Gusynin:2016som,Kotikov:2016prf}.

\begin{table}[t]
\centering
\begin{tabular}{|c||c|c|c|c|c|}
\hline
loops & 1 & 2 & 3 & 4 & 5 \\
\hline
$\al_{c}(N=0)$ & 1.0472 & 1.4872 & 1.1322 & ~~ --- ~~ & 1.0941  \\
\hline
\end{tabular}
\caption{Critical couplings of quenched QED$_4$ computed from \eqref{gap-eqn:de} up to $5$ loops 
(the symbol ``---'' indicates that no physical solution is found).}
\label{tab:1}
\end{table}

In the case of QED$_{4}$ ($\ep_e = 0$), \eqref{alc:RQED4de:two-loop:quenched} yields: $\al_{c}(N=0) = 4\pi/(6+\sqrt{6}) =1.4872$,
which is a new result and deviates substantially from the one-loop one (about $40\%$ increase).  
Actually, we can go beyond two-loops in QED$_4$ and apply \eqref{gap-eqn:de} with state of the art results for $\gamma_m$ up to 5 loops \cite{Luthe:2016xec,Baikov:2017ujl}.\footnote{We used the results of the ancillary files of \cite{Luthe:2017ttc} and \cite{Chetyrkin:2017bjc}.}
Our results for quenched QED$_4$ are summarized in Tab.~\ref{tab:1}. Though we  do not find any physical solution at 4 loops, we observe that, beyond two-loop, the critical coupling decreases reaching 
$1.0941$ at five-loops which deviates by only $4\%$ from $\pi/3$. 
This new result is in quantitative agreement with other estimates from numerical solutions of SD equations in quenched QED$_4$ with various vertex ans\"atze, see, \eg, \cite{Bloch:1995dd,Kizilersu:2014ela}, that lead to $\alpha_c(N=0)\approx 0.93$ in the Landau gauge\footnote{In quenched QED$_4$, up to 5 loops, the ``good gauge'' is very close (at most 0.18) to the Landau gauge. It is therefore reasonable to compare our gauge invariant results to those results of the literature that were calculated in the Landau gauge.}, \ie, a $15\%$ deviation from our $1.0941$.

\subsection{Application to unquenched RQED$_{4,d_e}$}

\subsubsection{General remarks}

In order to be able to consider the case of unquenched RQED$_{4,d_e}$, we shall follow 
\cite{Gorbar:2001qt,Gorbar:2002iw,Khveshchenko:2008ye,2009PhRvB..79t5429L,Kotikov:2016yrn} and include a dynamical 
screening of the interaction. In the case of RQED$_{4,3}$, this can be done with the help of the random phase approximation (RPA)
that amounts to a simple redefinition of the coupling constant \cite{Gorbar:2001qt,Teber:2012de} 
because, for this ``standing'' model, the polarization operator is finite (see \eqref{Pi1:exp:RQED43} combined with \eqref{dressed:dperp:RQED43}).
This procedure resums exactly all the $N$-dependence in $\gamma_m$ allowing us to go beyond the quenched approximation \cite{Kotikov:2016yrn}.

Following the RQED$_{4,3}$ case \cite{Kotikov:2016yrn}, we shall proceed with a generalization to RQED$_{4,d_e}$ by enforcing 
the resummation of the $N$-dependence of $\gamma_m$ at each loop order in an effective coupling constant. As we shall see below, 
such a procedure does not correspond to RPA in the case of QED$_4$. Indeed, applying RPA to QED$_4$ within our formalism does 
not allow to fully resum the $N$-dependence in $\gamma_m$. This originates from the running of the coupling constant which arises from
the UV divergent polarization operator (see \eqref{Pi1:exp:QED4}) and renders the QED$_4$ case more subtle to treat than the RQED$_{4,3}$ one.
Enforcing the resummation of the $N$-dependence of $\gamma_m$ in QED$_4$ is a semi-phenomenological prescription. But, as we shall see, 
it will provide results that are in quantitative agreement with those obtained from numerical solutions of SD equations. Moreover,
this effective coupling method will allow us to compute a critical fermion flavour number, $N_c$, such that $\al_c \ra \infty$, \ie,
a clearly non-perturbative quantity that requires the resummation of the $N$-dependence of $\gamma_m$. 

Within our formalism, an interesting fact about studying QED$_4$ is that it allows to consider higher orders in the loop expansion which are presently inaccessible in RQED$_{4,3}$.
We already exploited this in the last paragraph related to the quenched case. As we saw there, in both cases of
RQED$_{4,3}$ and QED$_4$, there is no solution to the 2-loop gap equation for $N>0$. It turns out that, for QED$_4$, solutions
of \eqref{gap-eqn:de} appear at higher loops for non-zero $N$ values without any need to introduce an effective coupling constant. 
We do not find these results satisfactory but, for completeness, we summarize them in Tab.~\ref{tab:2} and discuss them briefly. 

First, as can be seen from Tab.~\ref{tab:2}, a parity effect is observed as no solution
is obtained at 2- and 4-loop for $N>0$. 
At 3- and 5-loops, we obtain solutions over a range of $N$-values smaller that some $N_{\text{max}}$ which is now larger than $1$. 
However, this $N_{\text{max}}$ is not related to $N_c$ and it's physical interpretation is not clear; it is
close to the $N_c$ found from the effective coupling approach at 3-loop but deviates substantially from it at 5-loop, see  Tab.~\ref{tab:3} for the results of the effective coupling approach. Moreover,
at 5-loop the critical coupling obtained this way decreases with increasing $N$ while the opposite behaviour is observed at 3-loop. Actually, on physical grounds we expect that $\al_c$ should increase with increasing $N$. This comes from the fact that screening increases with $N$ thus effectively weakening interaction effects which in turn requires a larger value of the coupling constant in order to dynamically generate a mass. 

The effective coupling approach seems to overcome these difficulties (as far as our semi-phenomenological approach can tell) 
and we shall therefore focus on it in the following. 

\begin{table}[t]
\centering
\begin{tabular}{|c||c|c|c|c|c|}
\hline
loops            &    1     &   2   &    3       &     4     &   5     \\
\hline
$N_{\text{max}}$ & $\infty$ & 0.45  & 5.4738     &     0     & 18.4653 \\
\hline
$\al_{c}(N=0)$ & 1.0472 & 1.4872    & 1.1322     & ~~ --- ~~ & 1.0941  \\
$\al_{c}(N=1)$ & 1.0472 & ~~ --- ~~ & 1.1635     & ~~ --- ~~ & 1.0703  \\
$\al_{c}(N=2)$ & 1.0472 & ~~ --- ~~ & 1.2302     & ~~ --- ~~ & 1.0629  \\
$\al_{c}(N=3)$ & 1.0472 & ~~ --- ~~ & 1.3536     & ~~ --- ~~ & 1.0601  \\
$\al_{c}(N=4)$ & 1.0472 & ~~ --- ~~ & 1.6037     & ~~ --- ~~ & 1.0570  \\
$\al_{c}(N=5)$ & 1.0472 & ~~ --- ~~ & 2.2669     & ~~ --- ~~ & 1.0510  \\
$\al_{c}(N=6)$ & 1.0472 & ~~ --- ~~ & ~~ --- ~~  & ~~ --- ~~ & 1.0414  \\
$\al_{c}(N=7)$ & 1.0472 & ~~ --- ~~ & ~~ --- ~~  & ~~ --- ~~ & 1.0281  \\
\hline
\end{tabular}
\caption{Critical couplings of unquenched QED$_4$ computed from \eqref{gap-eqn:de} up to $5$ loops (the symbol ``---'' indicates that no physical solution is found).}
\label{tab:2}
\end{table}

\subsubsection{Effective coupling constant method}  
\label{subsec:effective-coupling}

In order to include dynamical screening in RQED$_{4,d_e}$, we define the following effective (reduced) coupling constant:
\be
\bar{g}_r = \frac{\bar{\al}_r}{1 + \hat{\Pi}_{1\,\text{eff}}^{(4,d_e)}\,\bar{\al}_r}, \qquad
\bar{\al}_r = \frac{\bar{g}_r}{1 - \hat{\Pi}_{1\,\text{eff}}^{(4,d_e)}\,\bar{g}_r}\, , 
\label{effective-coupling}
\ee
where $\bar{g}_r = g_r /(4\pi)$ and all the $N$-dependence, \ie, the effect of fermion loops (or screening), is by definition in $\hat{\Pi}_{1\,\text{eff}}^{(4,d_e)}$. 
Substituting \eqref{effective-coupling} in 
the mass anomalous dimension, \eqref{gamma:m:improved}, and expanding the resulting expression up to second order in $\bar{g}_r$,
we require that $\hat{\Pi}_{1\,\text{eff}}^{(4,d_e)}$ cancels the two-loop $N$-dependent terms in \eqref{gamma:m:improved}. This yields:
\be
\hat{\Pi}_{1\,\text{eff}}^{(4,d_e)} = 4(1-z)N\,\frac{(1-\ep_e)\,K_1}{3-2\ep_e} + \frac{10zN}{9}\, ,
\label{Pi1:4,de}
\ee
together with
\be
\gamma_m = 4\bar{g}_r\,\frac{3-2\ep_e}{(2-\ep_e)(1-\ep_e)} + 8\bar{g}_r^2\,\frac{3-2\ep_e}{(2-\ep_e)^3(1-\ep_e)^2} + \Ord(\bar{g}_r)^3\, .
\label{gammam:effective}
\ee
From \eqref{Pi1:4,de}, in the case of RQED$_{4,3}$ ($z=0$ and $\ep_e \ra 1/2$), we recover the fact that: 
\be
\hat{\Pi}_{1\,\text{eff}}^{(4,3)} = \hat{\Pi}_1^{(4,3)}/2\, ,
\ee
where $\hat{\Pi}_1^{(4,3)}$ was defined in \eqref{Pi1:exp:RQED43}, \ie, the procedure exactly corresponds to RPA in agreement with \eqref{dressed:dperp:RQED43}. In the case of QED$_4$ ($z=1$ and $\ep_e \ra 0$), we find 
from \eqref{Pi1:4,de} that
\be 
\hat{\Pi}_{1\,\text{eff}}^{(4)} = \hat{\Pi}_1^{(4)}/2\, ,
\ee
where $\hat{\Pi}_1^{(4)}$ was defined in \eqref{Pi1:exp:QED4}. As we anticipated, in the case of QED$_4$, 
the procedure is not RPA (see \eqref{dressed:dperp:QED4}) as it corresponds to a resummation of half of an effective one-loop renormalized polarization operator (at $L_p=0$). 

We may now proceed on solving \eqref{gap-eqn:de} with \eqref{gammam:effective}  truncated at one-loop order. The effective critical coupling $g_c$ that we obtain is equal to the one-loop critical coupling given by \eqref{alc:RQED4de:one-loop}. With the help of \eqref{effective-coupling}, this result can be generalized to the unquenched case and yields:
\be
\al_{c}(N)  = \frac{\pi\,(2-\ep_e)(1-\ep_e)^2}{2(3 - 2\ep_e) - \hat{\Pi}_{1\,\text{eff}}^{(4,d_e)}\,(2-\ep_e)(1-\ep_e)^2/4}\, . 
\label{alc:RQED4de:one-loop:unquenched}
\ee
Note that we can deduce from \eqref{alc:RQED4de:one-loop:unquenched} a critical fermion flavour number, $N_c$, such that $\al_c \ra \infty$. Its
expression reads:
\be
N_{c} = \frac{6\,(3 - 2\ep_e)^2}{3(1-z)(2-\ep_e)(1-\ep_e)^3 K_1 + 5 z }\, .
\label{Nc:RQED4de:one-loop:unquenched}
\ee

We now apply our general one-loop results to specific cases starting from RQED$_{4,3}$ ($z = 0$ and $\ep_e\rightarrow1/2$). In this case, 
\eqref{alc:RQED4de:one-loop:unquenched} and \eqref{Nc:RQED4de:one-loop:unquenched} yield:
\begin{flalign}
\al_{c}(N) =\frac{12\pi}{128-3\pi^2N}, \qquad N_{c}=\frac{128}{3\pi^2}=4.3230\, ,
\label{alc+Nc:res:one-loop:RQED43}
\end{flalign}
and, for the range of allowed non-zero values of $N$, \eqref{alc+Nc:res:one-loop:RQED43} yields:
\begin{flalign}
\al_{c}(N=1)=0.3832, \quad \al_{c}(N=2)=0.5481, \quad \al_{c}(N=3)=0.9624, \quad \al_{c}(N=4)=3.9415\, , 
\label{alc:N:res:one-loop:QED43}
\end{flalign}
thereby recovering in a simple and straightforward way the results of \cite{Gorbar:2001qt}. Following \cite{Gorbar:2001qt}, the result $\al_c(N)$ in \eqref{alc+Nc:res:one-loop:RQED43} defines a critical line in the $(\al,N)$ plane that separates the broken and symmetric phases. Moreover, in this one-loop case, 
$N_c = 4.3230$ is equal to the LO gauge-invariant critical fermion flavour number
of large-$N$ QED$_3$ (see discussion below \eqref{alc:RQED4de:one-loop}). 

Similarly, in the case of QED$_{4}$ ($z=1$ and $\ep_e\rightarrow0$), \eqref{alc:RQED4de:one-loop:unquenched} and \eqref{Nc:RQED4de:one-loop:unquenched} yield:
\be
\al_{c}(N)=\frac{18\pi}{54-5N}, \qquad N_{c}=\frac{54}{5}=10.8\, ,
\label{alc+Nc:res:one-loop:QED4}
\ee
and, for the range of allowed values of $N$, \eqref{alc+Nc:res:one-loop:QED4} yields:
\begin{flalign}
& \al_{c}(N=1)=1.1541, \quad \al_{c}(N=2)=1.2852, \quad \al_{c}(N=3)=1.4500, \quad  \al_{c}(N=4)=1.6632, 
\nonum \\
& \al_{c}(N=5)=1.9500, \quad \al_{c}(N=6)=2.3562, \quad \al_{c}(N=7)=2.9763, \quad \al_{c}(N=8)=4.0392,
\nonum \\
& \al_{c}(N=9)=6.2832,\quad \al_{c}(N=10)=14.1372\, .
\label{alc:N:res:one-loop:QED4}
\end{flalign}
Note that the one-loop SD equation in the rainbow approximation and including the polarization operator in the LAK-approximation (following the work of Landau, Abrikosov and Khalatnikov \cite{Landau:1956zr}) was approximately solved in the Landau gauge in \cite{Gusynin:1989mc} with the results $\al_{c}(N=1)=1.95$ (see
also \cite{Bloch:1995dd} for a review of other results as well as discussions below). 
It deviates by about $69\%$ from our result, $\al_{c}(N=1)=1.1541$.

Next, in order to access the two-loop case, we solve \eqref{gap-eqn:de} with the full \eqref{gammam:effective} as the input.
This yields the critical coupling $g_c=\al_{c}(N=0)$ where $\al_{c}(N=0)$ is the quenched critical coupling given by 
\eqref{alc:RQED4de:two-loop:quenched}. With the help of \eqref{effective-coupling}, this result can be generalized to the unquenched case  and yields:
\begin{flalign}
\al_{c}(N)  = \frac{\pi\,(2-\ep_e)(1-\ep_e)^2}{3 - 2\ep_e + \sqrt{\Delta_{0}} - \hat{\Pi}_{1\,\text{eff}}^{(4,d_e)}\,(2-\ep_e)(1-\ep_e)^2/4}\, . 
\label{alc:RQED4de:two-loop:unquenched}
\end{flalign}
Proceeding as in the one-loop case, we deduce from \eqref{alc:RQED4de:two-loop:unquenched} a critical fermion flavour number, $N_c$, such that $\al_c \ra \infty$. Its
expression reads:
\be
N_{c} = \frac{3\,(3 - 2\ep_e)\,(3 - 2\ep_e + \sqrt{\Delta_0})}{3(1-z)(2-\ep_e)(1-\ep_e)^3 K_1 + 5 z }\, .
\label{Nc:RQED4de:two-loop:unquenched}
\ee

We now apply our general two-loop results to specific cases starting from RQED$_{4,3}$ ($z = 0$ and $\ep_e\rightarrow1/2$). In this case, 
\eqref{alc:RQED4de:two-loop:unquenched} and \eqref{Nc:RQED4de:two-loop:unquenched} yield:
\begin{flalign}
\al_{c}(N) =\frac{36\pi}{32(6+\sqrt{6})-9\pi^2N}, \qquad N_{c}=\frac{32}{9\pi^2}\left(6+\sqrt{6}\right)=3.0440\, ,
\label{alc+Nc:res:two-loop:RQED43}
\end{flalign}
and, for the range of allowed non-zero values of $N$, \eqref{alc+Nc:res:two-loop:RQED43} yields:
\begin{flalign}
\al_{c}(N=1)=0.6230, \qquad \al_{c}(N=2)=1.2196, \qquad \al_{c}(N=3)=28.967 \, ,
\label{alc:N:res:two-loop:RQED43}
\end{flalign}
thereby recovering in a simple and straightforward  way all the results of \cite{Kotikov:2016yrn}. As already noticed in \cite{Kotikov:2016yrn}, we see from \eqref{alc+Nc:res:two-loop:RQED43} that the 2-loop $N_c$ in RQED$_{4,3}$ is slightly higher than the corresponding one in NLO large-$N$ QED$_3$ (for which $N_{c}=2.8470$, see discussion below 
\eqref{alc:RQED4de:two-loop:quenched}).

Similarly, in the case of QED$_{4}$ ($z=1$ and $\ep_e\rightarrow0$), \eqref{alc:RQED4de:two-loop:unquenched} and \eqref{Nc:RQED4de:two-loop:unquenched} yield:
\be
\al_{c}(N)=\frac{36\pi}{9(6+\sqrt{6})-10N}, \qquad N_{c}=\frac{9}{10}\left(6+\sqrt{6}\right)=7.6045\, ,
\label{alc+Nc:res:two-loop:QED4}
\ee
and, for the range of allowed values of $N$, \eqref{alc+Nc:res:two-loop:QED4} yields:
\begin{flalign}
& \al_{c}(N=1)=1.7124, \quad \al_{c}(N=2)=2.0180, \quad \al_{c}(N=3)=2.4562, \quad  \al_{c}(N=4)=3.1376, 
\nonum \\
& \al_{c}(N=5)=4.3423, \quad \al_{c}(N=6)=7.0486, \quad \al_{c}(N=7)=18.7080 \, .
\label{alc:N:res:two-loop:QED4}
\end{flalign}

Thanks to the knowledge of $\gamma_m$ up to 5 loops in QED$_4$, we can extend the above method to higher orders. 
At each order, the effective dynamical coupling is determined in such a way that it resums all the $N$-dependence of $\gamma_m$.  
This allows us to deduce a critical flavour fermion number, $N_c$, and the critical values of the coupling, $\al_c(N)$ for $N<N_c$. 
Our results for unquenched QED$_4$ (as well as the results of RQED$_{4,3}$ for clarity) are summarized in Tab.~\ref{tab:3}. As in previous approaches, no physical solution is found at 4 loops (all solutions
are complex) and therefore we effectively have  $N_c=0$ in this case. 
At 1-, 2-, 3- and 5-loops, a finite non-zero value of $N_c$ is obtained which decreases with increasing loop order. 
In agreement with our physical intuition, at 1-, 2-, 3- and 5-loops the critical coupling is seen to increase with $N$ (because of the increase of screening) and diverges at $N_c$. 
Moreover, provided that the apparent decrease of $N_c$ with increasing loop order is smooth, we are able to extrapolate its value to infinite loop order using a simple decreasing exponential fit (discarding the 4-loop order). This yields:
\be
N_c=2.0663\, .
\label{QED4:Nc:infinitel}
\ee
We therefore conjecture that, in the non-perturbative regime (infinite loop order), the critical coupling is defined only up to $N=2$. Moreover, we can also infer that, $\alpha_c(N=2)$ may be extremely large since $N=2$ is very close to $N_c=2.0663$.
Unfortunately, the values of the critical coupling oscillate too much thus preventing us from attempting a similar fit.


\begin{table}[t]
\centering
\begin{tabular}{|c||c|c||c|c|c|c|c|c|c|}
\hline
Model           & \multicolumn{2}{c||}{QED$_{4,3}$} & \multicolumn{7}{c|}{QED$_4$}                           \\
\hline \hline
loops           & 1         & 2         & 1      & 2         & 3         & 4   & 5         & ... & $\infty$  \\
\hline
$N_c$           & 4.3230    & 3.0440    & 10.8   & 7.6045    & 5.3298    & 0   & 3.3954    & ... & 2.0663    \\
\hline
$\al_{c}(N=0)$  & 0.2945    & 0.4183    & 1.0472 & 1.4872    & 1.1322    & --- & 1.0941    &     & ??        \\
$\al_{c}(N=1)$  & 0.3832    & 0.6229    & 1.1541 & 1.7124    & 1.2353    & --- & 1.2221    &     & ??        \\
$\al_{c}(N=2)$  & 0.5481    & 1.2196    & 1.2852 & 2.0180    & 1.4424    & --- & 1.6492    &     & ??        \\
$\al_{c}(N=3)$  & 0.9624    & 28.967    & 1.4500 & 2.4562    & 1.8707    & --- & 4.5576    &     & ~~ --- ~~ \\
$\al_{c}(N=4)$  & 3.9415    & ~~ --- ~~ & 1.6632 & 3.1376    & 2.9997    & --- & ~~ --- ~~ & ... & ~~ --- ~~ \\
$\al_{c}(N=5)$  & ~~ --- ~~ & ~~ --- ~~ & 1.9500 & 4.3423    & 11.150    & --- & ~~ --- ~~ &     & ~~ --- ~~ \\
$\al_{c}(N=6)$  & ~~ --- ~~ & ~~ --- ~~ & 2.3562 & 7.0486    & ~~ --- ~~ & --- & ~~ --- ~~ &     & ~~ --- ~~ \\
$\al_{c}(N=7)$  & ~~ --- ~~ & ~~ --- ~~ & 2.9763 & 18.708    & ~~ --- ~~ & --- & ~~ --- ~~ &     & ~~ --- ~~ \\
$\al_{c}(N=8)$  & ~~ --- ~~ & ~~ --- ~~ & 4.0392 & ~~ --- ~~ & ~~ --- ~~ & --- & ~~ --- ~~ &     & ~~ --- ~~ \\
$\al_{c}(N=9)$  & ~~ --- ~~ & ~~ --- ~~ & 6.2832 & ~~ --- ~~ & ~~ --- ~~ & --- & ~~ --- ~~ &     & ~~ --- ~~ \\
$\al_{c}(N=10)$ & ~~ --- ~~ & ~~ --- ~~ & 14.137 & ~~ --- ~~ & ~~ --- ~~ & --- & ~~ --- ~~ &     & ~~ --- ~~ \\
\hline
\end{tabular}
\caption{Critical couplings of unquenched QED$_{4,3}$ and QED$_4$, computed from \eqref{gap-eqn:de} together with the 
effective dynamical coupling \eqref{effective-coupling} up to $5$ loops 
(the symbol ``---'' indicates that no physical solution is found and the symbol ``??'' indicates that we could not compute the corresponding value).}
\label{tab:3}
\end{table}

In order to numerically compare our results with the ones of the literature, we should first discuss some ``good gauge'' issues. Indeed, at a given loop order greater than 1, a numerical evaluation of the ``good gauge'' parameter, $\xi_g(\alpha_c,N)$, shows substantial deviations from the Landau gauge as $N$ increases and even a divergence as $N$ approaches $N_c$. At $N=1$, a comparison with the Landau gauge results is justified since the ``good gauge'' is reasonably small, $0.28<\xi_g<0.48$. On the other hand, at $N=2$, a comparison with the Feynman gauge becomes more relevant ($0.41<\xi_g<0.89$). However, most of the unquenched SD results found in the literature are derived only in the Landau gauge. For these reasons, we are restricted to a comparison of only our $N=1$ results with the $N=1$ results of the literature in the Landau gauge.

Taking as a reference the review part of \cite{Kizilersu:2014ela}, table III, a wide range of results have been found for $\alpha_c(N=1)$ in the Landau gauge over the last 30 years. Most of the first order approaches (that neglect one or more equations of the SD system) presented in  \cite{Kizilersu:2014ela}, lead to results that are far from the results of our paper, \eg, a relative difference of $60\%$ up to $101\%$ by comparing with our 5 loop approach. 
However, better agreements are found when comparing our results with more sophisticated numerical approaches, especially the results of Bloch \cite{Bloch:1995dd}, that solve the full system of one-loop like SD equations with various vertex ans\"atze (bare, Ball-Chiu and modified CP as referred to in \cite{Bloch:1995dd}). These various approaches lead to results that are very close to each other in the $N=1$ case. Of course, the non-perturbative nature of the vertex ans\"atze used in these approaches does not have (to the best of our knowledge) a clear correspondence in terms of loops (such as in our case). Nevertheless, from these results, we find deviations of $2\%$--$5\%$ with respect to our two-loop results and deviations of $31\%$--$42\%$ with respect to our 4 and 5-loop results. We also note a smaller deviation of $22\%$ upon comparing our 5-loop $N=1$ result with the most recent results of \cite{Akram:2012jq}.

\section{Conclusion}
\label{sec.conclusion}

In this paper we have computed the two-loop mass anomalous dimension of RQED$_{4,d_e}$ with $N$ flavours of four-component
Dirac fermions. Our main formulas, \eqref{gamma:psi+m:improved}, for $\gamma_m$ and also the field anomalous dimension $\gamma_\psi$,
take into account the non-commutativity of the $\ep_\gamma \ra 0$ and $\ep_e \ra 0$ limits and are valid in both cases of RQED$_{4,3}$ (which
applies to graphene at its IR Lorentz invariant fixed point) and QED$_4$. For RQED$_{4,3}$, our formula for $\gamma_m$, \eqref{gamma:m:improved:RQED43:all}, generalizes the
one obtained in \cite{DiPietro:2019hqe} to an arbitrary number of (2-component) fermion flavours but our interest was mostly focused on the parity-even case \eqref{gamma:m:improved:RQED43}. When the later is mapped to large-$N$ QED$_3$ \cite{Kotikov:2016yrn}, it
corresponds exactly to the result obtained by Gracey in \cite{Gracey:1993sn} thereby strengthening our result.

We then proceeded on studying the critical properties (dynamical generation of a parity-conserving fermion mass) of RQED$_{4,d_e}$ (both in the quenched and the unquenched cases) with the help of the gap equation \eqref{gap-equation} (and its solution \eqref{gap-eqn:de}).
The latter was derived in a semi-phenomenological way on the basis of the modern approach of Gusynin and Pyatkovskiy \cite{Gusynin:2016som}. It matches exactly the NLO gap equation of
large-$N$ QED$_3$ \cite{Gusynin:2016som,Kotikov:2016prf} as well of that of RQED$_{4,3}$ \cite{Kotikov:2016yrn} and its range of application was extended to cover the case
of QED$_4$. Its only input being the mass anomalous dimension, it is fully gauge invariant by construction and allows to derive gauge-invariant critical coupling constants
and critical fermion flavour numbers of RQED$_{4,d_e}$ models with the help of \eqref{gamma:m:improved}. We also underlined the fact that 
quantities derived with the help of this gauge-invariant method
do match (at one-loop from our SD analysis) with gauge-dependant computations evaluated in the ``good gauge'' (a gauge where the fermion anomalous dimension vanishes order by order in perturbation theory). In the case of RQED$_{4,3}$, we have explicitly checked that \eqref{gap-eqn:de} does allow us to recover
in a simple and straightforward way all the NLO results of \cite{Kotikov:2016yrn}. In the case of QED$_4$, we have first recovered the celebrated $\al_c = \pi/3$
(now exactly gauge-invariant, see \eqref{alc:RQED4de:one-loop} with $\ep_e=0$)
at LO in agreement with the result obtained from solving the SD equations in the ``good gauge'' (see \eqref{alc:fromSD} with $\ep_e=0$). We have then used state of the art expressions
for $\gamma_m$ up to 5-loops in QED$_4$, to access the critical properties of the model. The 
unquenched case was considered first and our results are summarized in
Tab.~\ref{tab:1}. The latter shows that, at 5-loops, our value for the critical coupling, $\al_{c}(N=0)=1.0941$, deviates by only $4\%$ from $\pi/3$ and by $15\%$ from the results of \cite{Bloch:1995dd,Kizilersu:2014ela} where numerical solution of SD equations in the Landau gauge with various vertex ans\"atze lead to $\alpha_c(N=0)\approx0.93$. The unquenched case is more subtle due to the running of the
QED$_4$ coupling constant (in contrast to RQED$_{4,3}$ which is a ``standing'' gauge theory). Nevertheless, a resummation of the $N$-dependence of $\gamma_m$ into an effective coupling constant
gave us access to $N_c$ which is such that $\al_c \ra \infty$ together with all values of $\al_c(N)$ for $N<N_c$. Our results are summarized in
Tab.~\ref{tab:3}. At 5 loops, we find that $N_c = 3.3954$ with $\al_{c}(N=1)=1.2221$. Comparing our results to 
the ones of the literature, we find deviations of $2\%$--$5\%$ of our two-loop results with respect to the results of 
\cite{Bloch:1995dd} and deviations of $31\%$--$42\%$ of our 4 and 5-loop results with respect to the results of 
\cite{Bloch:1995dd}. We also note an interestingly smaller deviation of $22\%$ upon comparing our 5-loop $N=1$ result with the most recent results of \cite{Akram:2012jq}. From our results, we could extrapolate $N_c$ to infinite loop order finding $N_c=2.0663$, see \eqref{QED4:Nc:infinitel}. This suggests that dynamical mass generation in QED$_4$ may take place only for $N \leq 2$ (and hence for a possibly very large value of $\al_c(N=2)$).

Our work has shown that \eqref{gap-equation} (and its solution \eqref{gap-eqn:de}) allow for a simple and straightforward study of the critical properties of RQED$_{4,d_e}$.
In the case of RQED$_{4,3}$, we were limited to two-loop order as, to the best of our knowledge, $\gamma_m$ is still unknown for this model at 3-loop and beyond. It would be very interesting
to compute $\gamma_m$ at 3-loop for RQED$_{4,3}$ and use it as an input to \eqref{gap-eqn:de}. It would also be very instructive to solve the NNLO SD equations for RQED$_{4,3}$
along the lines of \cite{Kotikov:2016prf} for the NLO case, in order to have a solid proof that \eqref{gap-equation} does hold at this order. But such analytic computations may be quite
tedious at this order and it remains to be seen if they can even be carried out (let's recall that it took approximately 30 years since the seminal work of Nash \cite{Nash:1989xx} for the gauge-invariant NLO calculation in large-$N$ QED$_3$ to be completed in \cite{Gusynin:2016som,Kotikov:2016prf}). In the case of QED$_4$, our straightforward solution of the LO SD equations with the method of Kotikov
\cite{Kotikov:1989nm,Kotikov:2011kg} is a good indication that this method might be extended to NLO along the lines of \cite{Kotikov:2016prf} (but now for a running theory in the unquenched case). We leave all these projects for future investigations.

\acknowledgments

We would like to warmly thank Valery Gusynin and Anatoly Kotikov  for stimulating discussions and comments on our work as well as Lorenzo Di Pietro and Jingxiang Wu for fruitful discussions on \cite{DiPietro:2019hqe}. We would also like to warmly thank Roman Lee for kindly providing us access to the most recent version of LiteRed.

\appendix

\section{Master integrals}
\label{App:masters}

Following standard notations and conventions, see, \eg, the review \cite{Kotikov:2018wxe}, the one-loop Euclidean massless propagator-type integral is defined as:
\be
J(d_e,p,\alpha,\beta) = \int \frac{[\D^{d_e} k]}{k^{2\alpha}(k-p)^{2\beta}} = \frac{(p^2)^{d_e/2-\alpha-\beta}}{(4\pi)^{d_e/2}}\,G(d_e,\alpha,\beta)\, ,
\label{App:1-loop:J}
\ee
where $[\D^{d_e} k] = \D^{d_e} k / (2\pi)^{d_e}$. The dimensionless function $G(d_e,\al,\beta)$ is known exactly for all values of the indices $\al$ and $\beta$ and has a simple expression in terms
of $\Gamma$-functions:
\be
G(d_e,\alpha,\beta) = \frac{a(\alpha)\,a(\beta)}{a(\alpha+\beta-d_e/2)}, \qquad a(\alpha) = \frac{\Gamma(d_e/2-\alpha)}{\Gamma(\alpha)}\, .
\label{App:1-loop:G}
\ee
Similarly, the two-loop Euclidean massless propagator-type integral is defined as:
\begin{align}
J(d_e,p,\alpha_1,\alpha_2,\alpha_3,\alpha_4,\alpha_5)  &= \int \frac{[\D^{d_e} k_1] [\D^{d_e} k_2]}{(p-k_1)^{2\alpha_1}(p-k_2)^{2\alpha_2}{k_2}^{2\alpha_3}{k_1}^{2\alpha_4}(k_1-k_2)^{2\alpha_5}}
\nonum \\
&= \frac{(p^2)^{d_e-\sum_{i=1}^5 \alpha_i}}{(4\pi)^{d_e}}\,G(d_e,\alpha_1,\alpha_2,\alpha_3,\alpha_4,\alpha_5)\, .
\label{App:2-loop:J+G}
\end{align}
The dimensionless function $G(d_e,\alpha_1,\alpha_2,\alpha_3,\alpha_4,\alpha_5)$ is presently unknown for arbitrary values of the indices $\al_i$ ($i=1-5$) but has a long history
(see more in \cite{Grozin:2012xi} and \cite{Kotikov:2018wxe}). As can be seen from our exact results below, the only complicated two-loop master integral is 
$G(d_e,1-\ep_e,1,1-\ep_e,1,1)$ with arbitrary indices $\al_1=\al_3=1-\ep_e$. In the case of QED$_4$ ($\ep_e=0$), it's expression is known for a long time \cite{Chetyrkin:1980pr,Vasiliev:1981dg,Tkachov:1981wb,Chetyrkin:1981qh}. 
For arbitrary $\ep_e$, it has been represented as a combination of two ${}_3F_2$-hypergeometric functions of argument $1$ in \cite{Kotikov:2013eha} following the general method
of \cite{Kotikov:1995cw} (but the corresponding $\ep_\gamma$-expansion was not provided). The case $\ep_e=1/2$, relevant to RQED$_{4,3}$, has been recently encountered in \cite{Pikelner:2020mga} and 
it's leading-order epsilon expansion was computed in that paper. From the results of \cite{Pikelner:2020mga}, it is straightforward to find that:
\be
G(3 - 2\ep_\gamma,1/2,1,1/2,1,1) = \frac{8}{3\pi}\,\left(8\,\mathcal{C}+24\, \mbox{Cl}_4\left(\frac{\pi}{2}\right)\right)+\Ord(\ep_\gamma)\, ,
\label{App:2-loop:complicatedG}
\ee
where $\mathcal{C}\approx0.91597$ is the Catalan constant and $\mbox{Cl}_s(\theta)$ is the Clausen function with $\mbox{Cl}_4(\pi/2)\approx0.98895$\,. 
As noticed in \cite{Kotikov:2013eha}, the finite diagram (\ref{App:2-loop:complicatedG}) enters $\Sigma_V$ at two-loop with a factor proportional to $\ep_\gamma$ (see the last term in 
\eqref{App:Sigma2Vc} below). Hence, it does not contribute to $\Sigma_{2V}$ in the limit $\ep_\gamma \ra 0$ relevant to RQED$_{4,d_e}$. On the other hand, as can be seen from the last term in
\eqref{App:Sigma2Sc} below, (\ref{App:2-loop:complicatedG}) does contribute to $\Sigma_S$ at two-loop order in the limit $\ep_\gamma \ra 0$. 
But it only affects it's finite part. Hence, it does not contribute to the two-loop $Z_m$ and corresponding anomalous dimension $\gamma_m$ that we focus on in the main text.
    
\section{Exact results}
\label{App:exact}

In this appendix, we provide exact results for all the diagrams that where considered in the main text. All calculations were done by hand and crossed checked with the help of LiteRed \cite{Lee:2012cn,Lee:2013mka}. 
These results will be expressed in terms of the one and two-loop massless propagator-type master integrals presented in \ref{App:masters}.

Except for $\Sigma_{1S}$, all one-loop diagrams were previously computed in \cite{Kotikov:2013eha}. The expression of $\Sigma_{1S}$ and, for completeness, the ones of 
$\Sigma_{1V}$ and $\Pi_1$ are given by:
\begin{subequations}
\label{App:Sigma1+Pi1}
\begin{flalign}
&\Sigma_{1S}(p^2) = \frac{e^2}{(4\pi)^{d_\gamma/2}}\, \left(\frac{\mu^2}{-p^2}\right)^{\ep_\gamma} \Gamma(1 - \ep_e)\,
\frac{d_\gamma + d_e - 2 - (d_\gamma - d_e - 2)\,\xi}{2}\,G(d_e, 1, 1 - \ep_e)\, ,
\label{App:Sigma1S} \\
&\Sigma_{1V}(p^2) = \frac{e^2}{(4\pi)^{d_\gamma/2}}\, \left(\frac{\mu^2}{-p^2}\right)^{\ep_\gamma} \Gamma(1 - \ep_e)\,
\frac{d_e-2}{2}\,\left( \frac{d_\gamma -d_e}{d_\gamma + d_e - 4} - \xi  \right)\,G(d_e, 1, 1 - \ep_e)\, ,
\label{App:Sigma1V} \\
&\Pi_1(q^2) = -2N\,e^2\,\mu^{2\ep_\gamma}\, \frac{(-q^2)^{d_e/2-2}}{(4\pi)^{d_e/2}}\,\frac{d_e-2}{d_e-1}\,G(d_e,1,1)\, .
\label{App:Pi1}
\end{flalign}
\end{subequations}

At two-loop, the individual diagrams contributing to $\Sigma_S$ read:
\begin{subequations}
\label{App:Sigma2:S:a+b+c}
\begin{flalign}
&\Sigma_{2Sa}(p^2) = -2N\frac{e^4}{(4\pi)^{d_\gamma}}\, \left(\frac{\mu^2}{-p^2}\right)^{2\ep_\gamma}\Gamma^2(1 - \ep_e)\,
\frac{(d_e-2)(2d_\gamma+d_e-8)}{2d_\gamma -d_e - 4}\,G(d_e, 1, 1)\,G(d_e, 1, \ep_\gamma - \ep_e)\, ,
\label{App:Sigma2Sa} \\
&\Sigma_{2Sb}(p^2) = \frac{e^4}{(4\pi)^{d_\gamma}}\, \left(\frac{\mu^2}{-p^2}\right)^{2\ep_\gamma}\Gamma^2(1 - \ep_e)\,
\frac{(d_\gamma-3) (2d_\gamma + d_e -8)(d_\gamma+d_e-2-(d_\gamma-d_e-2)\xi)}{2 (d_\gamma-4)(d_\gamma+d_e-6)(d_\gamma+d_e-4)}\,
\nonum \\
&\hspace{2.2cm} \times \,\bigg( d_\gamma(d_\gamma-10) + d_e(4d_\gamma-d_e-2)+8 - (d_\gamma+d_e-4)(d_\gamma+d_e-6)\,\xi \bigg) \,
\nonum \\
&\hspace{2.2cm} \times \,G(d_e, 1, 1 - \ep_e)\,G(d_e, 1 - \ep_e, \ep_\gamma )\, ,
\label{App:Sigma2Sb} \\
&\Sigma_{2Sc}(p^2) = -\frac{e^4}{(4\pi)^{d_\gamma}}\, \left(\frac{\mu^2}{-p^2}\right)^{2\ep_\gamma}\Gamma^2(1 - \ep_e)\,\frac{d_e-2}{2}\, \Bigg\{
\left[  d_\gamma + 3 - \frac{3d_e}{2} + \frac{2d_e(d_e - 2)}{d_\gamma + d_e - 4} + \frac{2d_e}{2d_\gamma+d_e - 10} \right . \bigg .
\nonum \\
&\hspace{2.2cm} \left . - \left( 2d_\gamma - 2 - \frac{d_e (d_\gamma - d_e)}{d_\gamma + d_e -4} \right)\,\xi + \frac{2d_\gamma - d_e -6}{2}\,\xi^2 \right]\,
G(d_e,1, 1 - \ep_e)^2
\nonum \\
& \hspace{2.2cm} - \frac{2(d_\gamma-3)}{d_e-2}\,
\left[ d_\gamma - 4 + \frac{9d_e}{2} + \frac{2(d_e(d_e - 2)(d_e+4)+8)}{(d_e-2)(d_\gamma - 4)}
+ \frac{2(d_e - 4)(d_e(d_e-2)+4)}{(d_e-2)(d_\gamma + d_e - 6)}  \right .
\nonum \\
&\hspace{2.2cm} \left . - \frac{2(d_e - 2)^2}{d_\gamma + d_e - 4} -
\frac{(2d_\gamma + d_e -8)(d_e(3d_\gamma-4)+d_\gamma(d_\gamma-8)+8)}{(d_\gamma-4)(d_\gamma+d_e-4)}\,\xi +\frac{2d_\gamma + d_e -8}{2}\,\xi^2
\right]\,
\nonum \\
&\hspace{2.2cm} \times \, G(d_e, 1, 1 - \ep_e)\,G(1 - \ep_e, \ep_\gamma)
\nonum \\
& \hspace{2.2cm} \Bigg . - \frac{4(d_\gamma - 5) - d_e(d_e + 2(d_\gamma - 7))}{2d_\gamma + d_e - 10}\,
G(d_e,1-\ep_e,1,1-\ep_e,1,1) \Bigg \}\,.
\label{App:Sigma2Sc}
\end{flalign}
\end{subequations}
For completeness and ease of comparison, we also provide the expressions for the individual two-loop diagrams contributing to $\Sigma_V$ \cite{Kotikov:2013eha}:
\begin{subequations}
\label{App:Sigma2:V:a+b+c}
\begin{flalign}
&\Sigma_{2Va}(p^2) = 2N\,\frac{e^4}{(4\pi)^{d_\gamma}} \left(\frac{\mu^2}{-p^2}\right)^{2\ep_\gamma}\Gamma^2(1 - \ep_e)\,
\frac{(d_e-2)^2}{2d_\gamma -d_e - 6}\,G(d_e, 1, 1)\,G(d_e, 1, \ep_\gamma - \ep_e)\, ,
\label{App:Sigma2Va} \\
&\Sigma_{2Vb}(p^2) = \frac{e^4}{(4\pi)^{d_\gamma}} \left(\frac{\mu^2}{-p^2}\right)^{2\ep_\gamma}\Gamma^2(1 - \ep_e)\,
\frac{(d_e-2) (d_\gamma-3) (d_\gamma + d_e -4)}{2 (d_\gamma-4)}\,\left( \xi - \frac{d_\gamma - d_e}{d_\gamma + d_e -4} \right)^2 \,
\nonum \\
&\hspace{2.2cm} \times \,G(d_e, 1, 1 - \ep_e)\,G(d_e, 1 - \ep_e, \ep_\gamma )\, ,
\label{App:Sigma2Vb} \\
&\Sigma_{2Vc}(p^2) = -\frac{e^4}{(4\pi)^{d_\gamma}} \left(\frac{\mu^2}{-p^2}\right)^{2\ep_\gamma}\Gamma^2(1 - \ep_e)\,\frac{d_e-2}{2}\,  \Bigg\{ 
\left[  d_e - 4 + \frac{(d_e - 2)(d_\gamma - 3d_e + 4)}{2(d_\gamma + d_e - 4)} \right . \bigg . 
\nonum \\
&\hspace{2.2cm} - \frac{(d_\gamma + d_e - 6)(d_\gamma (d_e - 4) + 8)}{(2d_\gamma + d_e - 10)(2d_\gamma + d_e - 8)} 
- \frac{4(d_\gamma - d_e)}{d_\gamma + d_e - 4} - \frac{d_\gamma - d_e}{2d_\gamma + d_e - 8}\, 
\nonum \\
&\hspace{2.2cm} \times \left . \,\left(d_e - 8 - \frac{4(d_\gamma + d_e - 6)}{d_\gamma + d_e - 4} \right) 
- \xi\,\frac{(d_e - 2)(d_\gamma - d_e)}{d_\gamma + d_e - 4} + \xi^2\frac{d_e - 2}{2} \right]\,G(d_e,1, 1 - \ep_e)^2 
\nonum \\
& \hspace{2.2cm} + 
\left[ 2 d_e - d_\gamma - 1 + \frac{4(d_e - 2)(d_\gamma - 1)}{d_\gamma + d_e - 4} 
+ \frac{8(d_\gamma - 1)}{d_\gamma - 4} + \frac{2(d_e - 8)(d_\gamma - d_e)}{d_\gamma + d_e - 6} \right . 
\nonum \\
&\hspace{2.2cm} \left . - \frac{4(d_\gamma - 2)(d_\gamma - d_e)}{(d_\gamma - 4)(d_\gamma + d_e - 4)} 
+  2\xi\,\frac{(d_\gamma - 3)(d_\gamma - d_e)}{d_\gamma + d_e - 4} - \xi^2\,(d_\gamma - 3) \right]\,
\nonum \\
&\hspace{2.2cm} \times \, G(d_e, 1, 1 - \ep_e)\,G(d_e, 1 - \ep_e, \ep_\gamma) 
\nonum \\
& \hspace{2.2cm} \Bigg . - \frac{(d_\gamma - 4)(d_\gamma (d_e - 4) + 8)}{(2d_\gamma + d_e - 8)(2d_\gamma + d_e - 10)}\,
G(d_e,1-\ep_e,1,1-\ep_e,1,1) \Bigg \}\,.
\label{App:Sigma2Vc}
\end{flalign}
\end{subequations}

\bibliography{rqedm}



\end{fmffile}
\end{document}